\documentclass[submission,copyright,creativecommons]{eptcs}
\providecommand{\event}{VPT 2020} 
\usepackage{breakurl}             
\usepackage{underscore}           
\usepackage[latin1]{inputenc}
\usepackage[T1]{fontenc}
\usepackage[english]{babel}
\usepackage{amsmath}
\usepackage{amsfonts}
\usepackage{amssymb}
\usepackage{breakcites}
\usepackage{lmodern}     
\usepackage{array}
\usepackage{xspace}
\usepackage{enumitem}
\usepackage{hyperref}
\usepackage{vaucanson-g} 

\newcommand{\If}{\leftarrow}
\newcommand{\sm}[1]{{\small{$\mathtt{#1}$}}}
\newcommand{\U}{\raisebox{1.pt}{\bf \b{ }}}
\newcommand{\nil}{[\,]}  
\newcommand{\AX}{[A|X]}
\newcommand{\AY}{[A|Y]}
\newcommand{\AZ}{[A|Z]}
\newcommand{\AlY}{[A1|Y]}
\newcommand{\AsZ}{[A2|Z]}
\newcommand{\BZ}{[B|Z]}

\newcommand{\feq}{$\begin{array}{rlll}}
\newcommand{\nex}{\end{array}$\par $\begin{array}{rlll}}
\newcommand{\eeq}{\end{array}$}

\newcommand{\num}[1]{\hspace*{-.8cm}\makebox[.8cm][r]{#1}&}
\newcommand{\Feq}[1]{\feq\num{#1}}
\newcommand{\Nex}[1]{\nex\num{#1}}

\newcommand{\bpr}{\par\smallskip}
\newcommand{\epr}{\par\smallskip}

\title{A Historical Account of My Early\\ 
Research Interests\,\thanks{This work has been partially supported by GNCS-INdAM, 
Italy.}}

\author{Alberto~Pettorossi
\institute{DICII, University of Rome Tor Vergata, Rome, Italy}
\institute{IASI-CNR, Rome, Italy}
\email{adp@iasi.cnr.it}
}
\def\titlerunning{A Historical Account of My Early Research Interests}
\def\authorrunning{A. Pettorossi}	
	
\begin{document}
\maketitle

\begin{abstract}
This paper presents a brief account of some of the my early 
research interests.
This historical account starts from my laurea thesis on 
Signal Theory and my master thesis on
Computation Theory. It recalls some results in 
Combinatory Logic and Term Rewriting Systems. Some other results
concern Program Transformation,  Parallel Computation,
Theory of Concurrency, and Proof of Program Properties. 
My early research activity has been mainly done  in cooperation with
Andrzej Skowron, Anna Labella, and 
Maurizio Proietti.

\end{abstract}

\section{From Signal Theory to Combinatory Logic} 
\label{sec:SignalTheory-CombinatoryLogic}
\label{sec:SignTheory-TheoryofComp}

Since my childhood I very much liked Arithmetic and Mathematics. 
The formal reasoning always attracted my spirit and I always 
felt a special interest for numbers and geometrical
patterns. Maybe this was due to the fact that I thought that
Mathematics is a way of 
establishing `truth beyond any doubt'. As Plato says: 
`Truth becomes manifest in the mathematical process' (Phaedo). 
(The actual word used by Plato for `mathematical process' comes from 
$\lambda o 
\gamma \acute{\iota} \zeta o \mu \alpha \iota$  which means: I compute, I 
deduce.)
During my high school I attended the Classical Lyceum. 
Perhaps, for me the Scientific Lyceum would have been a better school to attend, but 
the Scientific Lyceum was located too far away from my home town.

At the age of nineteen, I began my university studies in 
Rome as a student of Engineering. I was in doubt whether or not 
to enrol myself as a  Mathematics student, but eventually I followed 
my father's suggestion to study Engineering because, as he
said: ``If you study Mathematics, you will have no other choice in life 
than to become a teacher.''
My thesis work was in Telecommunication and, in particular, I studied the problem 
of how to pre-distort an electric signal which encodes a sequence of
symbols, each one being 0 or 1,
via a sequence of impulses. The pre-distortion of the
electric signal should minimize the effect  of a Gaussian white 
noise (which would require a reduction of bandwidth) and the interference 
between symbols (which would require an increase of bandwidth).
A theoretical solution to this problem is not easy to find. Thus, 
I was suggested to look for a practical solution via a numerical simulation of the 
transmission channel and the construction of the so called {\it eye pattern}~\cite{Pet71}. 
In the numerical simulation, which uses the Fast Fourier Transform algorithm, 
one could easily modify the various parameters of 
the pre-distortion for minimizing  the errors in the  
output sequence of 0's and 1's. The thesis work was done under the 
patient guidance of my supervisors Professors Bruno Peroni and 
Paolo Mandarini.

After getting the laurea degree, I attended during 1972 at Rome University  
a course in Engineering
of Control and Computation Systems. During that year I read the book entitled
{\em Mathematical Theory of Computation} written by Professor Zohar 
Manna~(1939--2018)
(at that time that book was nothing more than a thick
technical report of Stanford University, California). I wrote my master 
thesis on the ``Automatic Derivation of Control Flow Graphs of Fortran Programs'',
under the guidance of Professor Vincenzo Falzone and Professor Paolo Ercoli~\cite{Pet72}. 
In particular, I wrote a Fortran program which
derives control flow graphs of Fortran programs. That program ran on a UNIVAC~1108 
computer with the EXEC~8 Operating System. The main memory had 128k words.
The program I wrote 
was a bit naive, but at that time I was not familiar with efficient
parsing techniques. I also studied various kinds of program schemas and, in 
particular,
those introduced by Lavrov~\cite{Lav61}, Yanov~\cite{Yan60}, and 
Martynuk~\cite{Mar65}.
Having constructed the control flow graph of a given Fortran program, one 
could transform that program into an equivalent one 
with better computational properties (such as smaller time or space 
complexities) 
by applying a set of schema transformations~\cite{ItZ71} which are guaranteed 
to preserve semantical equivalence. Schema transformations  
are part of the
research area in which I have been interested for some years afterwards.

During that period, which overlapped with my military service in the 
Italian Air Force, I also read a book on Combinatory Logic (actually, not
the entire book) by J.~R.~Hindley, B.~Lercher and J.~P.~Seldin~\cite{Hi&75,HiS86}.
I read the Italian edition of the book, which was emended of some 
inaccuracies with respect to the 
previous English edition (as Roger Hindley himself told me later).
Under the guidance of Professor Giorgio Ausiello and the great help of
my colleague Carlo Batini, I studied various properties 
of subbases in Weak Combinatory Logic (WCL)~\cite{BaP75}. 

WCL is an applicative system
whose terms, called {\it combinators}, can be defined as follows: (i)~$K$ and $S$ are atomic terms, and 
(ii)~if $t_{1}$ and $t_{2}$ are terms, then $(t_{1}\ t_{2})$ is a term.
When parentheses are missing, left associativity is assumed.
A notion of {\it reduction}, denoted~$>$, is introduced as follows:
for all terms $x, y, z$, $Sxyz > xz(yz)$ and
$Kxy > x$. Thus, for instance, $SKKS > KS(KS) > S$.
WCL is a Turing complete system as every
partial recursive function can be represented as a combinator in WCL.
A {\it subbase} in WCL is a set of terms which can be constructed starting a 
fixed set of (possibly non-atomic) combinators. For instance, the subbase~$\{B\}$, 
where~$B$ is a combinator defined
by the following reduction: $Bxyz > x(yz)$, is made out of all
terms which are constructed by $B$'s (and parentheses) only.
These terms are called $B$-combinators. One can show that $B$ can be expressed 
in the subbase $\{S,K\}$ by $S(KS)K$. Indeed, $S(KS)Kxyz >^{*} x(yz)$,
where $>^{*}$ denotes the reflexive, transitive closure of~$>$. 
The various subbases provide a way of partitioning the set of computable functions 
into various sets, according to the features of the combinators 
in the subbases. This should be contrasted with other stratifications
of the set of computable functions one could define and, among them,
the stratifications based on complexity classes or on
the Chomsky hierarchy~\cite{HoU79} with the 
{\it type~$i$} (for $i\!=\!0,1,2,3$) classes of languages.

Among other subbases, we studied  the subbase~$\{B\}$ and 
we showed how to construct the shortest \mbox{$B$-combinator} for  
constructing bracketed terms out of sequences of atomic subterms. For instance,
$B(B(BB)B)(BB)$ is the shortest $B$-combinator~$X$ such that:
$Xx_{1} x_{2} x_{3} x_{4} x_{5} x_{6} \ >^{*} \ x_{1} (x_{2} (x_{3} x_{4})) (x_{5} x_{6})$.

\smallskip
During 1975, while attending in Rome the conference on $\lambda$-calculus and 
Computer Science Theory, where our results on subbases 
were presented~\cite{BaP75}, 
I heard from Professor Henk Barendregt of an open problem concerning the
existence of a combinator~$\widetilde X$ made of only $S$'s (and parentheses), having 
no weak normal form. 
A combinator $T$ is said to be {\it in weak normal
form} if no combinator $T'$ exists such that $T\!>\!T'$. 
$X$ is said to have {\it weak normal form} if there exists
a combinator~$T$ such that $X >^{*} T$ and $T$ is in weak normal 
form. 

It was not hard to show that one such combinator $\widetilde X$ is 
$SAA(SAA)$, where $A$ denotes $((SS)S)$. 
I send the result
to Henk Barendregt (by surface mail, of course). Some years later
I was happy to see
 that an exercise about that problem and its solution 
was included in Barendregt's book on 
$\lambda$-calculus~\cite[page 162]{Bar84}.

\section{Finite and Infinite Computations} 
\label{sec:Finite_InfiniteComputation}
While studying Combinatory Logic, I became interested in terms viewed as 
trees and tree transformers.
Indeed, combinators can be considered both as 
trees and tree transformers at the
same time. This area was also related to the research on 
Term Rewriting Systems 
which was going to be one of my interests for a few years later.
The search for a non-terminating combinator stimulated my studies on
infinite, non-terminating computations.

In 1979 I 
introduced a hierarchy of infinite computations within WCL (and other
Turing complete systems)
which is related to the Chomsky hierarchy of languages~\cite{Pet79}.
That definition uses the notion of a {\em sampling function} $s$ which is
a total function from the set of 
natural numbers to $\{\mathit{true}, \mathit{false}\}$,
which from an infinite sequence $\sigma \!=\!\langle w_{0},w_{1},w_{2},
\ldots \rangle$ 
of finite words constructed by an infinite computation, 
selects an infinite subsequence $\sigma\!_{s}$ whose words are the elements of 
a (finite or infinite) language $L_{s}$. We state that 
$L_{s}=_{\mathit{def}} \{w_{j} \mid j\!\geq\!0 \wedge w_{j} ~\mathit{occurs~in}~
\sigma \wedge s(j)\!=\!\mathit{true}\}$. Let us assume that $L_{s}$ is 
generated by a grammar $G_{s}$.
In this case we say that also the subsequence $\sigma\!_{s}$ is generated by the
grammar~$G_{s}$.
Given a sequence~$\sigma$, by varying the sampling function~$s$ we have
different languages $L_{s}$ and different generating grammars $G_{s}$. For $i\!=\!0,1,2,3$, we say that the 
infinite computation which generates~$\sigma$ {\em is of type~$i$} 
if there exists a sampling function $s$ selecting a subsequence~$\sigma\!_{s}$ 
generated by a grammar of type~$i$, and no sampling function~$s'$
exists such that the subsequence selected by $s'$ is generated by a grammar of type~$(i\!+\!1)$.

For instance, let us consider the following program $P$\,:\nopagebreak

\makebox[84mm][l]{$w=``a$''$\,;~~~ {\mathtt{while}}  ~true~ {\mathtt{do}}~  
{\mathit{print~w\,;}}~~ w = ``b$''$\,w\,``c$''$\,;~~ \pi_{0}~{\mathtt{od}}$}

\noindent 
where $a,b,c$ are characters, $w$ is a string of characters, and 
$\pi_{0}$ is a terminating program fragment associated with a type~0 language $L_{0}$,
such that: (i)~$L_{0}$ is not of type 1, (ii)~$\pi_{0}$ does not modify~$w$,
(iii)~at each loop body execution, $\pi_{0}$ prints only one word of $L_{0}$,
 and
(iv)~for every word $v\!\in\! L_{0}$ there is exactly one body execution in
which
$\pi_{0}$ prints $v$.
We have that $P$ evokes an infinite computation of type~2, as the grammar with
axiom $S$ and productions: $S\rightarrow  a \mid b\,S\,c$~ is a type~2 (context free) grammar.

%

When I first presented this hierarchy definition at a conference, I met my dear colleague Philippe 
Flajolet (1948-2011) and he said to me: ``I have already studied these 
topics~\cite{FlS74}. You 
should look at the immune sets.'' 
That remark motivated my first encounter with Roger's book on 
recursivity~\cite{Rog67} where immune sets are defined and analyzed.
Then also Professor Maurice Nivat (1937-2017) came to me and said: ``It is 
a nice piece of work,... 
but you should rewrite the paper in a better way!''. I was very glad that Nivat showed interest 
in my work. 
He was right in asking me to rewrite it and improve it.
Unfortunately, I did not follow his suggestion. Not even when, 
a few years later, 
Professor Tony Hoare told me: ``I like writing and rewriting my papers.'' 

Looking for terms with infinite behaviour in WCL, in 1980 
I wrote a paper on the automatic construction of combinators having 
no normal form
by using the so called {\em accumulation} 
method and the {\em pattern matching and hereditary embedding} 
method~\cite{Pet80a}. 
The solutions of some equations between terms would guarantee the existence 
of the combinators with the desired properties. 

On the other side of the camp, that is, considering the finite behaviours, many people at that time were studying properties
of Term Rewriting Systems (TRSs)  which would guarantee termination. Among them,
Nachum Dershowitz, Samuel Kamin, Jean-Jacques L\'evy, and David Plaisted. 
In 1981 I wrote a paper 
introducing the {\it non-ascending property}~\cite{Pet81a}. In that paper
I related the various techniques which were proposed,
including recursive path orderings, simplification ordering, 
and bounded lexicographic orderings. 
I thank Nachum for pointing out to me some errors
in that paper and, in particular, a missing left-linearity 
hypothesis about the TRS under consideration~\cite{Der87}. 
A TRS is said to be 
{\em left linear} if the variable occurrences on the left hand side of every 
rule are all distinct. For instance, 
$f(x,y,z) \rightarrow f(y,z,x)$ is a left linear rule, while $f(x,y,x)
 \rightarrow g(y,x)$
is not. During a conference coffee-break,
Jean-Jacques showed me a simple inductive proof of 
Fact~1~\cite[pages~436--437]{Pet81a} 
using bounded lexicographic orderings (actually, that proof is based on a 
non-predicative definition of the non-ascending rewriting rules).

\section{Program Transformation} 
\label{sec:ProTrans}

During the years 1977--1981 I visited Edinburgh University. 
I was supported by the 
British Council organization and the Italian National Research Council. I did my Ph.D.
thesis work on program transformation under the guidance of Professor Rod
Burstall 
and also Professor Robin Milner, 
during Rod's visit to Inria in Paris for some months. 
I met Rod in person for the first time at the Artificial Intelligence Department, in 
Hope Park Square at Edinburgh. 
I addressed him by saying: ``Professor Burstall,$\ldots$''. I do not remember 
my subsequent words, but I do remember what he said to me in answering:
``Alberto, this is the last time you call me `professor'. Please, call me Rod.''
He introduced me to functional programming and he wrote `for~me', as he said,
a compiler for a new functional language, called NPL~\cite{Bur77}
he was developing at that time. The language
NPL later evolved into Hope~\cite{Bu&80}. While at Edinburgh, 
I wrote a paper~\cite{Pet78} on the 
automatic annotation of functional programs for improving memory utilization. 
Functions could 
destroy the value of their arguments whenever they were no longer needed for
subsequent computations. I apologize for not having Rod as
co-author of that paper. 

My Ph.D.~thesis work was mainly on program transformation 
starting from the seminal paper by Rod and John Darlington~\cite{BuD77}. 
Some time before, Rod had received a letter from Professor Edger W.~Dijkstra 
(1930-2002) proposing the following `exercise' 
in program transformation:
the derivation of an iterative program for the {\tt fusc} 
function~\cite[pages~215--216, 230--232]{Dij82}:

{\small{

\makebox[36mm][l]{${\mathtt{fusc(0) = 0}}$} ${\mathtt{fusc(1) = 1}}$

\makebox[36mm][l]{${\mathtt{fusc(2n) = fusc(n)}}$} ${\mathtt{fusc(2n\!+\!1) = fusc(n\!+\!1) + fusc(n)}}$  ~~~~for  ${\mathtt{n\!\geq\!0 }}$
}}


\noindent
In one of my scientific conversations with Rod,
 he told me about his research interests and he
also mentioned the above exercise. 
The difficult part of the exercise was how to 
motivate the `invention' of the new function definitions
to be introduced during program transformation 
in the so called {\it eureka steps}~\cite{BuD77}. 

To do the same exercise Bauer and W\"ossner~\cite[page 288]{BaW82} 
use an embedding into a linear combination, that is, they define the function 
{\small{${\mathtt{F(n,a,b)=_{\mathit{def}} }}$}}  
\sm{a\! \times\! {\mathtt{fusc}}(n) + b \!\times \! fusc(n\!+\!1)}.
Using that function, they are able to derive 
for {\small{\tt{fusc}}} a program that is linear recursive and also 
tail-recursive. Then, from that program 
they easily derive an iterative program.
But, where the function {\tt F} comes from? I wanted to do the exercise
using the unfolding/folding rules only~\cite{BuD77} and, at the same time,
I wanted to
give a somewhat mechanizable account of the definition the
new functions to be introduced. 

Now, the unfolding rule allows one to
unroll (upto a specified depth) the recursive calls thereby generating a directed acyclic graph of 
distinct calls. I called that graph the {\it m-dag}.
The prefix {\it m} (short for minimal) tells us that in an  {\it m-dag} 
identical function calls are denoted by a single node.  
Then, I used the so called {\it tupling strategy} that allows one to 
define new functions as the result of tupling 
together function calls which share
common subcalls, that is, calls which have common descendants in the m-dag.
Note that to check this sharing property requires syntactic 
operations only on the m-dags.
By using the {tupling strategy}, looking at the m-dag for \sm{fusc},
we introduce the tuple function \sm{t(n)=_{def}\langle fusc(n),~ fusc(n\!+\!
1)\rangle} and we get the following recursive equations for \sm{fusc}:

\vspace{2mm}
{\small{

\makebox[28mm][l]{${\mathtt{fusc(n) = u}}$} ${\mathtt{where~ \langle u,v \rangle = t(n)}}$   
~~for  ${\mathtt{n\!\geq\!0 }}$ 

\vspace{1mm} 

\makebox[9mm][l]{${\mathtt{t(0) =}}$} \{by unfolding\} \sm{=\langle fusc(0),~fusc(1)\rangle =} \{by unfolding\}  \sm{= \langle 0,1\rangle}

\makebox[12mm][l]{${\mathtt{t(2n) =}}$}\{by unfolding\} \sm{=\langle fusc(2n),\ fusc(2n\!+\!1)\rangle =} \{by unfolding\} =\\
\makebox[14mm][l]{}\sm{=~\langle fusc(n),\ fusc(n\!+\!1)\!+\!fusc(n)\rangle =} \{by \sm{where} abstraction~\cite{BuD77}\} =\\  
\makebox[38mm][l]{\makebox[14mm][l]{}\sm{=~ \langle u,u\!+\!v\rangle}}\sm{where \langle u,v\rangle = 
\langle fusc(n),~fusc(n\!+\!1)\rangle} =  \{by folding\} =\\
\makebox[38mm][l]{\makebox[14mm][l]{}{${\mathtt{=~ \langle u, u\!+\!v\rangle}}$}}{\tt{where}} ${\mathtt{\langle u, v\rangle = t(n)}}$ ~~for  ${\mathtt{n\!>\!0}}$


\makebox[31mm][l]{${\mathtt{t(2n\!+\!1)\! =\! \langle u\!+\!v, v\rangle}}$} {\tt{where}} ${\mathtt{\langle u, v\rangle = t(n)}}$ ~~for  ${\mathtt{n\!\geq\!0}}$ \hfill(by a 
derivation similar to that of \sm{t(2n)})~~  
}}

\vspace{2mm}
\noindent
Now a last step is needed to get the iterative program desired by 
Dijkstra's exercise. 

I used the following schema equivalence (such as the ones in~\cite{WaS73})
stating that \sm{t(n)} defined by the non-tail recursive equations:

\vspace{1mm}
\sm{t(0)=a}

\makebox[40mm][l]{\sm{t(2n)= b(t(n))}}   for \sm{n\!>\! 0}

\makebox[40mm][l]{\sm{t(2n+1)= c(t(n))}} for \sm{n\!\geq\! 0}

\vspace{1mm}
\noindent 
is equal to the value of \sm{res} returned by the following program, 
where ${\mathtt{B[\ell..0]}}$ stores the 
binary expansion of~\sm{m}, the most significant bit being at position $\ell$ 
(obviously, 
\sm{B[\ell..0]}  can be computed by performing $O(\log$ \sm{m}$)$
successive integer divisions by~2):

\vspace{1mm}
{\small{
\makebox[26mm][l]{${\mathtt{res \!=\! a;~~~ p\!=\!\ell;}}$}
${\mathtt{while~ p\!\geq\! 0~ do~ if~~ B[p]\!=\!0 ~~then~~ res \!=\! b(res)
~~else~~ res\! =\! c(res)\,; ~~p = p\!-\!1 ~~od}}$
}}

\vspace{1mm}
\noindent
By using this schema equivalence we derive from the above linear, non-tail 
recursive program for \sm{fusc} the following 
iterative program:

\vspace{2mm}
{\small{
\makebox[70mm][c]{${\mathtt{\{n\!\geq\! 0}} ~~\wedge~$ \sm{n= \sum_{p=0}^{\ell} B[p]\cdot 2^{p}}\}}


\vspace{.5mm} 
${\mathtt{\langle u,v \rangle  =  \langle 0,1 \rangle;}}$
~~~~ ${\mathtt{p  = \ell;}}$\rule{0mm}{3.5mm}\nopagebreak
\vspace{1mm}

\makebox[18mm][l]{${\mathtt{while~ p\!\geq\! 0}}$}
\makebox[5mm][l]{${\mathtt{do~}}$}${\mathtt{if~~ B[p]\!=\!0 ~~then~~ v=u\!+\!v ~~else~~ u=u\!+\!v\,;}}$\makebox[4mm][l]{}${\mathtt{p = p\!-\!1 ~od}}$

\vspace{1mm}
\makebox[70mm][c]{$\{{\mathtt{\langle u,v\rangle\! =\! t(n)}} ~~\wedge~~ {\mathtt{u \!=\! fusc(n)}}\}$}
}}

\vspace{1mm}
\noindent 
Note that we do not need to state the somewhat intricate invariant 
of the while-loop
for showing the correctness of the derived iterative program, as Dijkstra's 
methodology for program
construction would have required us to do. The derived program, which is
correct by construction, uses
an $O(\log$ \sm{n}$)$ number of operations for computing \sm{fusc(n)}
as Dijkstra's program reported
in~\cite[page 215--216]{Dij82}\footnote{In order to get exactly Dijkstra's program, 
one should perform a generalization step as indicated
in~\cite{Pet84a}.}\!.
We have only to show by induction, once and for all, the validity of the schema 
equivalence we have used.

Having derived an iterative program for the {\tt fusc} function,
I faced the problem of deriving by transformation an iterative program,
such as the one suggested by~\cite{MiB66}, which computes the Fibonacci function
 \sm{fib(n)} using an $O(\log$\,{\sm{n}}$)$ number
of arithmetic operations. Here is the definition
of the Fibonacci function:


{\small{
${\mathtt{fib(0)\!=\!0  \hspace{20mm} fib(1)\!=\!1}}$   

${\mathtt{fib(n\!+\!2) = fib(n\!+\!1) + fib(n)}}$  
~for  ${\mathtt{n\!\geq\!0}}$    \hfill$(\dagger1)$~
}}

\noindent
By using the tupling strategy  the function 
{\small{${\mathtt{g(n) =_{\mathtt{def}} \langle  fib(n),\ fib(n\!-\!1) 
\rangle}}$}} is introduced and the following program is derived:

\vspace{1mm}

{\small{
${\mathtt{fib(0)\!=\!0  \hspace{20mm}  fib(1)\!=\!1}}$
  
${\mathtt{fib(n\!+\!2)\!=\!u}}$ ~~{\tt{where}} ${\mathtt{\langle u, v\rangle = g(n\!+\!2)}}$  
\hspace{10mm} for  ${\mathtt{n\!\geq\!0}}$

\makebox[9mm][l]{${\mathtt{g(1) =}}$} 
\sm{\langle 1,0\rangle}


\makebox[28mm][l]{${\mathtt{g(n\!+\!2)\!=\!\langle u\!+\!v, u\rangle}}$} {\tt{where}} 
${\mathtt{\langle u, v\rangle = g(n\!+\!1)}}$  \hspace{5mm} for  ${\mathtt{n\!\geq\!0}}$
}}

\vspace{1mm}
\noindent
The iterative program for {\small{\tt fib}} can be obtained by applying 
the following schema equivalence stating that {\small{$\mathtt{g(n)}$}} defined 
by the  equations:

\vspace{1mm}
{\small{
\makebox[18mm][l]{${\mathtt{g(0)\!=\!a}}$}  ${\mathtt{g(n\!+\!1)\!=\!
b(g(n))}}$
 ~for  ${\mathtt{n\!\geq\!0}}$
}}

\vspace{1mm}
\noindent
is equal to the value of {\small{$\mathtt{res}$}} returned by the program:

{\small{
\makebox[18mm][l]{${\mathtt{res \!=\! a;}}$}
${\mathtt{while~ n\!>\! 0~ do~~ res = b(res)\,; ~~n = n\!-\!1 ~~od}}$
}}

\vspace{1mm}
\noindent 
Thus, we get:

{\small{
\makebox[42mm][c]{${\mathtt{\{n\!\geq\! 0}}\}$}
    
${\mathtt{if~ n\!=\!0 ~then~ u\!=\!0 ~else}}$\rule{0mm}{3.5mm}

${\mathtt{if~ n\!=\!1 ~then~ u\!=\!1 ~else}}$

${\mathtt{begin~ p= n\!-\!1;~ \langle u,v \rangle \!=\!\langle 1,0 \rangle;~ }} {\mathtt{~while ~ p\!>\!0 ~do~ \langle u,v \rangle\! =\!\langle u\!+\!v, u \rangle\,;~~ p = p\!-\!1 ~od ~end}}$\nopagebreak

\makebox[42mm][c]{$\{{\mathtt{u\! =\! fib(n)}}\}$}
}}

\noindent
This program has a linear time complexity, in the sense that it computes the result by 
a linear number of additions. In order to get a program which requires 
$O(\log$\,\sm{n}$)$ arithmetic operations when computing \sm{fib(n)},
we should invent the {\it multiplication}
operation, which is not present in Equation~\sm{(\dagger 1)}. From that equation by 
unfolding we have:

\vspace{1mm}
{\small{
\makebox[52mm][l]{\makebox[16mm][l]{${\mathtt{fib(n\!+\!2)}}$}${\mathtt{= fib(n\!+\!1) 
+ fib(n) =}}$}
\{by unfolding ${\mathtt{fib(n\!+\!1)}}$\}  =

\makebox[55mm][l]{\makebox[16mm][l]{}${\mathtt{= 2\cdot fib(n) + fib(n\!-\!1)=}}$}
\{by unfolding ${\mathtt{fib(n)}}$\}  =

\makebox[16mm][l]{}${\mathtt{= 3\cdot fib(n\!-\!1) + 2\cdot fib(n\!-\!2)}}$
\hfill$(\dagger2)$~
}}

\vspace{1mm}
\noindent
The unfolding process may continue for some more steps, but we stop here. 
We will not discuss here the important issue of how many unfolding steps
should be performed when deriving programs by transformation. 
Let us simply note that more unfoldings may exhibit more patterns 
of function calls from which more efficient functions can be derived.

In our case the invention of the multiplication 
operation is reduced to three generalization steps~\cite{PeB82}. 
First, we generalize the 
initial values {\small{\tt 0}} and {\small{\tt 1}} of the function 
{\small{${\mathtt{fib}}$}} to two variables
{\small{${\mathtt{a_{0}}}$}}  and {\small{${\mathtt{a_{1}}}$}}, respectively.
(This kind of generalization step 
is usually done when mechanically proving theorems about functions~\cite{BoM75}.) 
By promoting those new variables to arguments, we get the following new 
function~\sm{G}: 

\vspace{1mm}
{\small{${\mathtt{G(a_{0},\!a_{1},\!0)\!=\!a_{0}  \hspace{20mm} G(a_{0},\!a_{1},\!1)\!=\!
a_{1} }}$}}

{\small{${\mathtt{G(a_{0},\!a_{1},\!n\!+\!2) = G(a_{0},\!a_{1},\!n\!+\!1) +
G(a_{0},\!a_{1},\!n)}}$ \hspace{10mm} for ${\mathtt{n\!\geq\!0}}$   \hfill$(\dagger3)$~
}}

\vspace{1mm}
\noindent
This function~{\small{$\mathtt G$}} satisfies the following equation 
which is derived from Equation~{\small{$(\dagger3)$}}, as 
Equation~{\small{$(\dagger2)$}} has been derived from {\small{$(\dagger1)$}}:

\vspace{1mm}
 {\small{${\mathtt{G(a_{0},a_{1},n\!+\!2)\!=\!3\cdot 
G(a_{0},a_{1},n\!-\!1)+ 2\cdot G(a_{0},a_{1},n\!-\!2)}}$ \hfill$(\dagger4)$~ 
}}

\vspace{1mm}
\noindent
The second generalization consists in generalizing the coefficients \sm{2} 
and \sm{3} 
to two functions \sm{p(n)}  and \sm{q(n)}, 
respectively (and thus, multiplication is introduced). By this generalization we
establish a correspondence between the value of the coefficients and the
number of unfoldings performed. We can then derive the explicit definitions
of the functions \sm{p(n)}  and  \sm{q(n)} as shown in~\cite{PeB82}, and we get that
\sm{p(n)=G(1,0,n)}  and \sm{q(n)=G(0,1,n)}.

The third, final generalization consists in generalizing the argument 
\sm{n\!+\!2}
on the left hand side of Equation~{\small{$(\dagger4)$}} to 
{\small{${\mathtt{n\!+\!k}}$}} and promoting 
the new variable 
{\small{$\mathtt k$}} to an argument of a new function defined as follows:
{\small{$\mathtt{F(a_{0},a_{1},n,k) =_{def} G(a_{0},a_{1},n\!+\!k)}$}}.

From the equations defining {\small{$\mathtt{F(a_{0},a_{1},n,k)}$}}
we get (the details are in~\cite[pages 184--185]{PeB82}): 

\vspace{1mm}
 {\small{${\mathtt{G(a_{0},a_{1},n\!+\!k) = G(0,1,k) \cdot 
G(a_{0},a_{1},n\!+\!1)+G(1,0,k) \cdot G(a_{0},a_{1},n)}}$}}
\vspace{1mm}

\noindent
Then, by taking {\small{${\mathtt{n\!=\!k}}$}} and 
{\small{${\mathtt{n\!=\!k\!+\!1}}$}}, we also get:

\vspace{1mm}
\makebox[110mm][l]{{\small{${\mathtt{G(a_{0},a_{1},2k)= G(0,1,k) \cdot 
G(a_{0},a_{1},k\!+\!1)+G(1,0,k) \cdot G(a_{0},a_{1},k)}}$}}} for~\sm{k> 0}
\nopagebreak

\makebox[110mm][l]{{\small{${\mathtt{G(a_{0},a_{1},2k\!+\!1)= G(0,1,k) \cdot 
G(a_{0},a_{1},k\!+\!2)+G(1,0,k) \cdot G(a_{0},a_{1},k\!+\!1)}}$}}} for~\sm{k\geq 0}

\vspace{1mm}
\noindent
Eventually, by tupling
together the function calls which share the same subcalls, we get the following 
program which computes \sm{fib(n)} by performing an $O(\log$\,\sm{n}$)$ number of 
arithmetic operations only, as desired.
For all \sm{k\geq 0}, the function \sm{r(k)} is  the pair 
\sm{\langle G(1,0,k),~ G(0,1,k)\rangle}.

\vspace{1mm}
{\small{

\makebox[42mm][l]{${\mathtt{fib(0) = 0  \hspace{20mm}  fib(1)=1}}$}\nopagebreak 

${\mathtt{fib(n\!+\!2)\!=\!u\!+\!v ~~where~ \langle u,v \rangle = r(n\!+\!1)}}$   
~for  ${\mathtt{n\!\geq\!0 }}$\nopagebreak 

${\mathtt{r(0) = \langle 1,0\rangle}}$\nopagebreak

\makebox[54mm][l]{${\mathtt{r(2k) = \langle u^{2}\!+\!v^{2},\ 2uv\!+\!v^{2}\rangle}}$} {\tt{where}} ${\mathtt{\langle u, v\rangle = r(k)}}$ ~~for  ${\mathtt{k\!>\!0}}$ 
\nopagebreak   

\makebox[54mm][l]{${\mathtt{r(2k\!+\!1) = \langle 2uv\!+\!v^{2},\ (u\!+\!v)^{2}\!+\!v^{2}\rangle}}$} {\tt{where}} ${\mathtt{\langle u, v\rangle = r(k)}}$ ~~for  ${\mathtt{k\!\geq\!0}}$   
}}

\vspace{1mm}
\noindent
We leave to the reader to derive the iterative program that can be obtained 
by a simple schema equivalence from this program. 
One can say that the program we have derived is even better than 
the program based on $2\hspace{-.5mm}\times\hspace{-.7mm}2$ matrix multiplications~\cite{MiB66},
because it tuples together two values only, not four, as required by the use of
the $2\hspace{-.5mm}\times\hspace{-.7mm}2$ matrices. 
Note that our derivation of the program does not rely on any knowledge 
of matrix theory.

In the paper with Rod Burstall~\cite{PeB82} there is a generalization of the
derivation of the logarithmic time program for {\small{\tt fib}} 
to the case of any linear recurrence relation over any 
semiring structure. What remains to be done? One may want to derive a {\it constant time}
 program
for evaluating any linear recurrence relation over a semiring. This would require
the introduction of the {\it exponentiation} operation. Recall that 
{\small{${\mathtt{fib(n) = (A^{n}-B^{n})/sqrt(5)}}$}}, where 
{\small{${\mathtt{A =(1\!+\!sqrt(5))/2}}$}}
and  {\small{${\mathtt{B =(1\!-\!sqrt(5))/2}}$}}.

\smallskip
From September 1977 to June 1978, I visited  the School of Computer and Information Science at Syracuse 
University, N.Y., USA. I attended courses taught by Professor Alan Robinson,
John Reynolds, Lockwood Morris, and Robert Kowalski 
(at that time a visiting professor from Imperial College, London, UK).
It was a splendid occasion for deepening my knowledge about many aspects 
of Computer Science from such illustrious teachers. 

In Syracuse I had the opportunity of reading more carefully some parts 
of the book 
{\it Automata Theory, Languages, and Computation} by  Hopcroft and 
Ullman~\cite{HoU79} and the 
book {\it Introduction to Mathematical Logic} by Mendelson~\cite{Men87}. 
I was exposed
by Professor Kowalski for the first time to various topics of Artificial 
Intelligence and I read the 
preliminary draft of his beautiful book {\it Logic for Problem 
Solving}~\cite{Kow79b}. 
I remember the stress put by Kowalski on Keith 
Clark's
{\it negation as failure} semantics for logic programs~\cite{Cla78}.
This Computational Logic area was going to become my main research 
area in the years to come, through
my cooperation with Maurizio Proietti in Logic Program Transformation.

\section{The Tupling Strategy and the List Introduction Strategy} 
\label{sec:TuplingListIntro}
The results of the use of tupling and generalization during program transformation
were presented in a paper of the 1984 ACM Symposium on
Lisp and Functional Programming, Austin, Texas, USA~\cite{Pet84a}.
While giving a seminar on those results at the University of Warsaw (Poland)
Professor Helena Rasiowa\footnote{Helena Rasiowa and Roman Sikorski 
gave in 1950 a first algebraic proof of G\"odel Completeness Theorem for first-order predicate calculus.}
who was in the audience, at the end kindly said to me: ``Your paper is a 
collection of examples!''. I was not surprised by that remark,  
but I was happy to have, among the examples,
a simple derivation of an iterative program for computing 
the moves of the Towers of Hanoi problem. 
That task was considered to be very challenging by some authors (see, for 
instance,~\cite[page 285]{Hay77}), and the derivation I proposed is also
easily mechanizable. 

The following Hanoi function {\small{${\mathtt{h(n,A,B,C)}}$}}  computes 
the shortest sequence of moves in
the free monoid {\small{$\mathtt{\{AB,BC,CA,BA,CB,AC\}^{*}}$}} to move 
{\small{${\mathtt{n\,(\geq\!0)}}$}}
disks from peg~{\small{${\mathtt{A}}$}} to peg~{\small{${\mathtt{B}}$}} using 
peg~{\small{${\mathtt{C}}$}} as an extra peg. A move of a disk from peg 
{\small{\tt X}} 
to peg {\small{\tt Y}}  is denoted by~{\small{\tt XY}},  for any distinct 
{\small{\tt X}}, {\small{\tt Y}} in {\small{$\mathtt{\{A,B,C\}}$}}. Every disk is of a
different size
and over any disk only smaller disks can be placed. $\varepsilon$ denotes the 
empty sequence of moves, and 
{\small{${\mathtt{::}}$}} denotes the concatenation of sequences of moves.

\vspace{1mm}
{\small{
${\mathtt{h(0,A,B,C) = \varepsilon }}$

${\mathtt{h(n\!+\!1,A,B,C)=
h(n,A,C,B)::AB::h(n,C,B,A)}}$   
~~for  ${\mathtt{n\!\geq\!0 }}$   \hfill$(\dagger5)$~
}}

\begin{figure}[ht!]
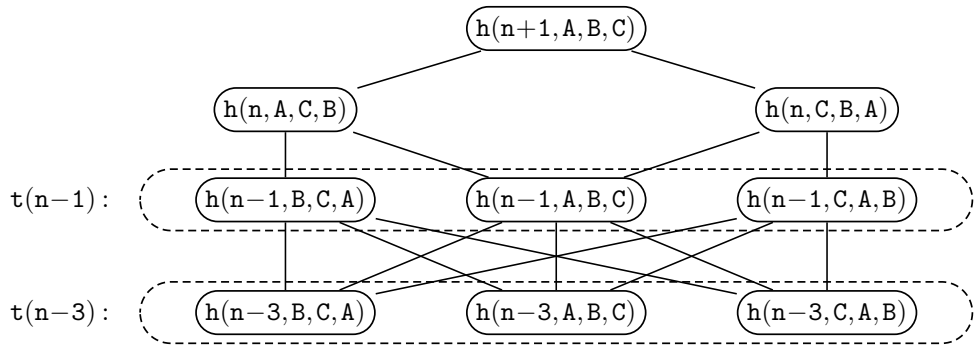

\begin{center}
\VCDraw{%
\begin{VCPicture}{(-4,-1)(14,6.8)} 
\SetEdgeArrowWidth{0pt}
\ChgStateLineWidth{.6}
\FixStateDiameter{10mm}

\StateVar[$\sm{h(n\!+\!1,A,B,C)}$]{(6,5.8)}{nABC}  
\StateVar[$\sm{h(n,A,C,B)}$]{(0,4)}{n1ACB} 
\StateVar[$\sm{h(n,C,B,A)}$]{(12,4)}{n1CBA} 

\FixStateDiameter{14mm}
\ChgStateLineWidth{.6}
\ChgStateLineStyle{dashed}
\StateVar[\hspace{106mm}]{(6,2)}{Tuple} 
\ChgStateLineColor{white}
\ChgStateLineStyle{solid}
\StateVar[$\sm{t(n\!-\!1):}$]{(-5,2)}{Tuple1} 
\ChgStateLineColor{black}

\ChgStateLineWidth{.6}
\FixStateDiameter{10mm}
\StateVar[$\sm{h(n\!-\!1,B,C,A)}$]{(0,2)}{n2BCA} 
\StateVar[$\sm{h(n\!-\!1,A,B,C)}$]{(6,2)}{n2ABC} 
\StateVar[$\sm{h(n\!-\!1,C,A,B)}$]{(12,2)}{n2CAB} 
\FixStateDiameter{14mm}
\ChgStateLineWidth{.6}
\ChgStateLineStyle{dashed}
\StateVar[\hspace{106mm}]{(6,-.5)}{Tuple} 

\ChgStateLineColor{white}
\ChgStateLineStyle{solid}
\StateVar[$\sm{t(n\!-\!3):}$]{(-5,-.5)}{Tuple3}  
\ChgStateLineColor{black}

\ChgStateLineWidth{.6}
\FixStateDiameter{10mm}
\StateVar[$\sm{h(n\!-\!3,B,C,A)}$]{(0,-.5)}{n3BCA} 
\StateVar[$\sm{h(n\!-\!3,A,B,C)}$]{(6,-.5)}{n3ABC} 
\StateVar[$\sm{h(n\!-\!3,C,A,B)}$]{(12,-.5)}{n3CAB} 

\EdgeL{nABC}{n1ACB}{} \EdgeL{nABC}{n1CBA}{} 
\EdgeL{n1ACB}{n2BCA}{} \EdgeL{n1ACB}{n2ABC}{} 
\EdgeL{n1CBA}{n2ABC}{} \EdgeL{n1CBA}{n2CAB}{}

\EdgeL{n2BCA}{n3BCA}{} \EdgeL{n2BCA}{n3ABC}{} \EdgeL{n2BCA}{n3CAB}{}
\EdgeL{n2ABC}{n3BCA}{} \EdgeL{n2ABC}{n3ABC}{} \EdgeL{n2ABC}{n3CAB}{}
\EdgeL{n2CAB}{n3BCA}{} \EdgeL{n2CAB}{n3ABC}{} \EdgeL{n2CAB}{n3CAB}{}
\end{VCPicture}
}
\end{center} \normalsize \caption{An upper portion of the call graph {\it m-dag} of the Hanoi
function \sm{h(n\!+\!1,A,B,C)}. An edge from an upper node to a lower node denotes that the 
upper call requires the lower call. Dashed lines denote tuples.\label{fig:mdag-hanoi}}
\vspace*{0mm}
\end{figure}

\vspace{1mm}
\noindent
In order to get an iterative program for computing {\small{${\mathtt{h(n,A,B,C)}}$}},
we first unfold {\small{${\mathtt{h(n,A,C,B)}}$}} and 
{\small{${\mathtt{h(n,C,B,A)}}$}} in {\small{$(\dagger5)$}} and then we tuple together 
in the new function {\small{$\mathtt{t(n\!-\!1)}$}} the calls of
{\small{$\mathtt{h(n\!-\!1,A,B,C)}$}}, {\small{$\mathtt{h(n\!-\!1,B,C,A)}$}}, and 
{\small{$\mathtt{h(n\!-\!1,C,A,B)}$}}
which share common subcalls (see Figure~\ref{fig:mdag-hanoi}). The order 
of the components in the tuple is insignificant. Details are 
in~\cite{Pet85a}.

We get:\nopagebreak

{\small{
\makebox[30mm][l]{${\mathtt{h(0,A,B,C) = \varepsilon}}$} 
${\mathtt{h(1,A,B,C) = AB}}$\nopagebreak

\makebox[68mm][l]{${\mathtt{h(n\!+\!2,A,B,C) = u::AC::v::AB::w::CB::u}}$}${\mathtt{where~
\langle u,v,w \rangle =t(n)}}$ ~~for  ${\mathtt{n\!\geq\!0 }}$\nopagebreak

\makebox[30mm][l]{${\mathtt{t(0) = \langle \varepsilon,\varepsilon,\varepsilon\rangle}}$}
${\mathtt{t(1) =  \langle AB,BC,CA \rangle}}$\nopagebreak

\makebox[17mm][l]{${\mathtt{t(n\!+\!2) = \langle}}$}${\mathtt{u::AC::v::AB::w::CB::u,}}$ \makebox[3mm][l]{}${\mathtt{v::BA::w::BC::u::AC::v,}}$\nopagebreak


\makebox[68mm][l]{\makebox[17.5mm][l]{}${\mathtt{w::CB::u::CA::v::BA::w \, \rangle}}$}${\mathtt{where~ \langle u,v,w \rangle =t(n)}}$ 
~~for  ${\mathtt{n\!\geq\!0 }}$ 
}}

\vspace{1mm}
\noindent
Then, we can apply the schema equivalence stating that 
{\small{$\mathtt{g(n)}$}} defined 
by the  equations:

\vspace{1mm}
{\small{
\makebox[18mm][l]{${\mathtt{g(0)\!=\!a}}$}  ~~~~~ ${\mathtt{g(1)\!=\!b}}$ 
\hspace{15mm} ${\mathtt{g(n\!+\!2)\!=\!c(g (n))}}$
 ~~~for  ${\mathtt{n\!\geq\!0}}$
}}

\vspace{1mm}
\noindent
is equal to the value of {\small{$\mathtt{res}$}} returned by the program:

\vspace{1mm}
{\small{
\makebox[18mm][l]{${\mathtt{if~ even(n) ~~then~~ res \!=\! a ~~else~~ res \!=\! b;}}$}

${\mathtt{while~ n\!>\! 1~ do~~ res = c(res)\,; ~~n = n\!-\!2 ~~od}}$
}}

\vspace{1mm}
\noindent
We get the following program, where  for {\small{$\mathtt{k\!=\!1,2,3}$}}, {\small{$\mathtt{Tk}$}} denotes the {\small{$\mathtt{k}$}}-th component
of the triple~\sm{T}\,: 

\vspace{1mm}
{\small{
\makebox[46mm][c]{${\mathtt{\{n\!\geq\! 0}}\}$}
    
\vspace{1mm}
${\mathtt{if~~ n\!=\!0 ~~then~~ Hanoi\!=\!\varepsilon ~~~else}}$\rule{0mm}{3.5mm}

${\mathtt{if~~ n\!=\!1 ~~then~~ Hanoi\!=\! AB ~~else}}$\rule{0mm}{3.5mm}

${\mathtt{begin~~ n\!=\!n\!-\!2; ~~if~ even(n) ~~then~~ T\!=\!\langle
\varepsilon,  \varepsilon, \varepsilon \rangle ~~else~~ T\!=\!\langle
AB,  BC, CA \rangle\,; }}$

\makebox[36mm][l]{\hspace*{4mm}${\mathtt{while~ n\!>\! 1 
~do~~ T\! =\!\langle\,}}$}${\mathtt{T1\!::\!AC\!::\!T2\!::\!AB\!::\!T3\!::\!CB\!::\!T1,}}$
\makebox[2mm][l]{}${\mathtt{T2\!::\!BA\!::\!T3\!::\!BC\!::\!T1\!::\!AC\!::\!T2,}}$


\makebox[36mm][l]{}${\mathtt{T3\!::\!CB\!::\!T1\!::\!CA\!::\!T2\!::\!BA\!::\!T3\, \rangle; ~~~~ n\!=\!n\!-\!2 ~~od\,;}}$

\hspace*{4mm}\sm{Hanoi=T1\!::\!AC\!::\!T2\!::\!AB\!::\!T3\!::\!CB\!::\!T1}

\sm{end}

\makebox[60mm][c]{$\{\,{\mathtt{Hanoi = h(n,A,B,C)}}\,\}$}
}}

\vspace{1mm}
\noindent
The technique we have presented is based only on the tupling strategy and a simple 
schema equivalence. That technique is successful also 
for the many variants of the Towers of Hanoi problem that can be found in the literature
(see, among others,~\cite{Er83}). 
A different derivation for computing the Hanoi function can be done by
introducing, besides the tuple \sm{t(n)}, also the tuple 
\sm{t'(n\!-\!2)=_{def} \langle h(n\!-\!2,A,C,B),} \sm{h(n\!-\!2,C,B,A),} 
\sm{h(n\!-\!2,B,A,C) \rangle} corresponding to the calls of \sm{h} 
at level \sm{n\!-\!2} (not depicted in Figure~\ref{fig:mdag-hanoi}). 
We leave this derivation to the reader.

\vspace{1mm}
In a later paper I addressed the problem of finding the \sm{m}-th move of 
algorithms which
compute sequences of moves without computing any other move~\cite{Pet87a}. 
This problem arose as a generalization of the
problem relative to the Towers of Hanoi. If the moves are computed by a function
defined by a recurrence relation, then under suitable hypotheses, it is indeed 
possible to compute the \sm{m}-th move without computing any other move.
For the case of the Hanoi
function \sm{h(n,A,B,C)} we have that the length \sm{Lh(n)} of the sequence of 
moves 
for \sm{n} disks, satisfies the following equations:
{\small{
${\mathtt{Lh(0) = 0  \hspace{10mm} Lh(n\!+\!1)=
2 \cdot Lh(n) \!+\! 1}}$   
~~for  ${\mathtt{n\!\geq\!0 }}$   
}}

\begin{figure}[h!]
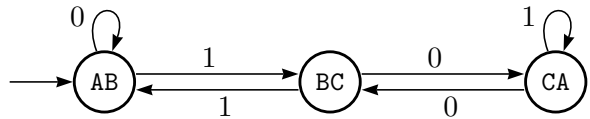

\begin{center}
\VCDraw{%
\begin{VCPicture}{(-2,-.8)(12,2)} 
\SetEdgeArrowWidth{7pt}
\SetEdgeArrowLengthCoef{1.5}
\ChgStateLineWidth{1.}
\FixStateDiameter{14mm}

\State[$\sm{AB}$]{(0,0)}{AB} 
\State[$\sm{BC}$]{(5,0)}{BC} 
\State[$\sm{CA}$]{(10,0)}{CA} 

\Initial{AB} 

\ForthBackOffset
\SetLoopAngle{20} \SetLoopOffset{2pt}
\EdgeL{AB}{BC}{1} \EdgeL{BC}{AB}{1} \LoopN{AB}{0} 
\EdgeL{BC}{CA}{0} \EdgeL{CA}{BC}{0} \LoopN{CA}{1} 
\end{VCPicture}
}
\end{center} \normalsize \caption{The finite automaton for computing the 
\sm{m}-th move in the sequence \sm{h(n,A,B,C)} of moves 
for the Towers of 
Hanoi problem with \sm{n} disks and pegs \sm{A}, \sm{B}, and \sm{C}.
\label{fig:fa-hanoi}}
\end{figure}

One can show~\cite{Pet87a} that the \sm{m}-th move of \sm{h(n,A,B,C)}, 
for \sm{1\!\leq\! m \!\leq \!2^{n}\!-\!1} and \sm{n\!\geq\!0},
can be computed using the 
deterministic finite automaton of Figure~\ref{fig:fa-hanoi}. We assume that
\sm{M[\ell..0]} is the binary expansion of \sm{m}, the most significant bit being at 
the leftmost position \sm{\ell}. Thus, \sm{m=\sum_{i=0}^{\ell} M[i]\cdot 2^{i}} and
\sm{m} is not a power of \sm{2} iff \sm{M[\ell..0]\not \in 10^{*}}.
Let \sm{trans(X,p)} denotes the state \sm{Y} such that in the finite automaton of 
Figure~\ref{fig:fa-hanoi} there is an arc  from state \sm{X} to state 
\sm{Y} with label~\sm{p}.

\vspace{2mm}
{\small{

\sm{i\!=\!\ell; ~~state\!=\!AB;}

\sm{while~M[i..0] \not \in 10^{*}  ~do~}
\sm{begin~~ state = trans(state, M[i]); ~~~i=i\!-\!1~~ end ~od}
\rule{0mm}{4.mm}

}}

\vspace{2mm}
\noindent
The \sm{m}-th move is the name of the final state, with \sm{B} and \sm{C} interchanged if 
an odd number of state transitions is made. 

Suppose that we want to compute the \sm{44}-th move of \sm{h(6,A,B,C)}. The binary expansion of 
\sm{44}  is \sm{101100}. Starting from the left, we take the prefix \sm{101} up to (and excluding) the suffix in \sm{10^{*}} (in our case \sm{100}). 
We perform the transitions on the automaton of Figure~\ref{fig:fa-hanoi} 
starting from state \sm{AB} 
according to that prefix (from left to right) and we get to state~\sm{CA}. 
Since the length of the prefix
 is odd (it is indeed \sm{3}), the move to be computed is \sm{BA}, that is, \sm{CA} with \sm{B} and \sm{C} interchanged.

\smallskip
In a subsequent paper with Maurizio Proietti~\cite{PeP02b} we want to explore 
the idea of introducing
{\it lists}, rather than {\it arrays} (indeed, tuples being of fixed size can be seen as arrays). 
Originally, this idea was suggested to me by Rod Burstall.  
Since every recursive function 
can be computed by using stacks (actually, two stacks are sufficient for computing any
partial recursive function on natural numbers~\cite{HoU79}), this technique seems to me, at first, not 
very relevant in the practice of improving the time complexity of a program or avoiding
inefficient recursions. We explored the use of this technique and, indeed, we managed to 
achieve good results. In particular, the {\it list introduction strategy}
can be used when the recursive calls do not
generate a sequence of {\em cuts} of {\em constant} size in the m-dag of the
function calls, and thus it does not allow the use of the tupling strategy. 
A cut in an m-dag is set $C$ of nodes such that every path from the root to a leaf
intersects $C$. In the case of the Hanoi function 
(see Figure~\ref{fig:mdag-hanoi}) we have depicted 
the cuts associated with \sm{t(n\!-\!1)} and 
\sm{t(n\!-\!3)}. Both of them are of size~\sm{3} and thus,
the tupling strategy (with three function calls) is successful. 
More details on cuts and their use for program transformation
also in relation with pebble games~\cite{PaH70} can be found in 
my Ph.D.~thesis~\cite{Pet84c}.

We used the list introduction strategy for deriving a program for computing 
the binomial coefficients:
$\binom{n+1}{k+1}\!=\! \binom{n}{k}\! +\! \binom{n}{k+1}$. In this case the 
sequence of cuts
from the root to the leaves is of increasing size. Indeed, 
$\binom{n+1}{k+1}$ requires the computations of
$\binom{n}{k}$ and $\binom{n}{k+1}$, which in turns, require the  computations 
of $\binom{n-1}{k-1}$, $\binom{n-1}{k}$,  $\binom{n-1}{k+1}$, and so on. (Indeed, 
in the Pascal Triangle the basis has an increasing size when the height 
of the triangle increases). Therefore, the tupling strategy cannot be used.

Now, in order to show the power of the list introduction strategy, let us consider 
the \mbox{$n$-queens} problem. Details are in~\cite{PeP02b}. 
An $n\!\times\!n$ board configuration {\it Qs} is
represented by a list of pairs of the form:
$[\langle R_1, C_1 \rangle, \ldots, \langle R_n, C_n \rangle]$, 
where for $i\! =\! 1,\ldots,n$,  $\langle R_i, C_i \rangle$
denotes a queen placed in row~$R_i$ and column~$C_i$. For $i \! =\! 1,\ldots,n$,
the values of $R_i$ and~$C_i$ belong to the list $[1,\ldots,n]$.
 
We start from the following initial program {\it Queens\/}:

\noindent
\bpr
\Feq{1.} {\it queens}({\it Ns}, {\it Qs}) \If
            {\it placequeens}({\it Ns}, {\it Qs}),
          \ {\it safeboard}({\it Qs})
\Nex{2.} {\it placequeens}([\,], [\,]) \If
\Nex{3.} {\it placequeens}({\it Ns}, [Q|{\it Qs}]) \If
          {\it select}(Q, {\it Ns}, {\it Ns}1),\
          {\it placequeens}({\it Ns}1,{\it Qs})
\Nex{4.} {\it safeboard}([\,]) \If
\Nex{5.} {\it safeboard}([Q| {\it Qs}]) \If
          {\it safequeen}(Q, {\it Qs}),\ {\it safeboard}({\it Qs})
\Nex{6.} {\it safequeen}(Q, [\,]) \If
\Nex{7.} {\it safequeen}(Q1, [Q2| {\it Qs}])
          \If {\it notattack}(Q1, Q2),\ {\it safequeen}(Q1, {\it Qs})
\eeq
\epr

\noindent
In order to place 
$n$ queens we solve the goal
${\it queens}([1,\ldots,n],{\it Qs\/})$.
By clause~1 we have that 
${\it placequeens}([1,\ldots,n], {\it Qs})$ generates 
a board configuration {\it Qs} and ${\it safeboard}({\it Qs})$
checks that in~{\it Qs\/} no
two queens lie on the same diagonal (either `up diagonal' or 
`down diagonal' in Dijkstra's terminology~\cite{Dij71}).
We assume that 
 ${\it notattack\/}(Q1, Q2)$ holds iff queen position (or
queen, for short) $Q1$, that is, $\langle R1, C1 \rangle$, is not on the same
diagonal of the queen $Q2$. The tests that the queens
are neither on the same row nor on the same column can be avoided
by assuming that
${\it select\/}(Q, {\it Ns}, {\it Ns}1)$ holds
iff ${\it Ns}$ is a
list of distinct numbers in $[1,\ldots,n]$,
$Q$ is queen~$\langle R,C \rangle$ such that
row  $R$ is the length of {\it Ns} and
column $C$ is a member of {\it Ns},
and ${\it Ns}1$ is the new list
obtained from ${\it Ns}$ by deleting the occurrence of~$C$. 
The length of the list ${\it Ns}$
decreases by one unit after each call of {\it placequeens}.
In particular, we have that board configurations
having~$k$~queens (with $1\! \leq\! k\! \leq\! n$) are of the form:
$[\langle n, c_1 \rangle, \langle n\!-\!1, c_2 \rangle,
\ldots, \langle n\!-\!k\!+\!1, c_k \rangle]$,
where $c_1, c_2, \ldots, c_k$ are distinct numbers in
$[1, \ldots, n]$.

Program {\it Queens} solves the problem using the {\it generate-and-test\/} 
approach and it is not efficient. A more
efficient program using an {\em accumulator} that stores the diagonals which
are not safe, 
has been proposed in~\cite[page 255]{StS94}. Efficiency is increased because 
backtracking is reduced. 

By applying the list introduction strategy (which includes also some 
generalization steps)
one can derive the following program {\it TransfQueens} whose behaviour is
similar to that of the accumulator version. The various transformation steps are described in~\cite{PeP02b}.
The higher efficiency of the final program is
due to the fact that the test for a safe board configuration is `promoted' into
the process of generating new configurations, and the number of generated
unsafe board configurations is decreased (see the
{\it filter promotion\/} technique~\cite{Bir84b,Dar78}).

\nopagebreak
\bpr 
\Feq{8.}{\it queens}([\,],[\,]) \If
\Nex{9.}{\it queens}({\it Ns},[Q |{\it Qs}])\If 
  		{\it select\/}(Q, {\it Ns}, {\it Ns}1),\
		{\it genlist}1({\it Ns}1,{\it Qs},[Q])
\Nex{10.}{\it genlist}1([\,],[\,],{\it Ps}) \If 
\Nex{11.}{\it genlist}1({\it Ns},[Q1 |{\it Qs}],[\,]) \If
        	{\it select}(Q1,{\it Ns},{\it Ns}1),\
		{\it genlist\/}1({\it Ns}1,{\it Qs},[Q1])
\Nex{12.}{\it genlist}1({\it Ns},[Q1 |{\it Qs}],[P1|{\it Ps}]) \If
        	{\it select}(Q1,{\it Ns},{\it Ns}1),\  {\it notattack}(P1,Q1),

		\nex\hspace{58.5mm}{\it genlist\/}2({\it Ns}1,{\it Qs},[P1],{\it Ps},Q1)
\Nex{13.}{\it genlist}2({\it Ns}1,{\it Qs},{\it Ps}1,[\,],Q1) \If
        	{\it genlist\/}1({\it Ns}1,{\it Qs},[Q1|{\it Ps}1])
\Nex{14.}{\it genlist}2({\it Ns}1,{\it Qs},{\it Ps}1,[P2|{\it Ps}2],Q1) \If 
        	{\it notattack}(P2,Q1),\\
	
    \nex\hspace{68mm}{\it genlist\/}2({\it Ns}1,{\it Qs},[P2|{\it Ps}1],{\it Ps}2,Q1)
\eeq
\epr

\noindent
This  program performs much less backtracking than the {\em Queens}
program\footnote{In some experiments we have done, for 10 queens {\it TransfQueens} runs about  
70 times faster than {\em Queens}.}\!. 
By clause~9, the first queen position $Q$ is selected
and {\it genlist\/}1 is called with its
last argument storing the current board configuration,
which is the list $[Q]$. 
When a new queen is placed at position $Q1$ and
it is not attacked by the last queen placed at position $P1$
(see clause~12), {\it genlist\/}2 
checks whether or not $Q1$ is attacked by the queens already present 
in the current configuration and whose positions are stored in {\it Ps\/}
(see clauses~13 and 14).
If $Q1$ is not attacked,
the  configuration is updated  (see the last argument $[Q1|Ps1]$ of 
{\it genlist\/}1 in clause 13).
Otherwise, if~$Q1$ is attacked, 
by backtracking (see the atom {\it select} in clause~12), a
different queen position is selected.
If all positions for the new queen are under attack, then by backtracking
(see the atoms {\it select}  in clauses~9 and 11), the
position of a previously placed queen, if there is one, is selected in a 
different way.

The explanation which we have just given about the derived program 
(clauses~8--14),  
may appear unclear to the non-expert reader, but one should note that 
it was not needed at all. Indeed, correctness of the derived program is 
guaranteed by the correctness of the transformation rules, and the 
efficiency improvement 
is due to filter promotion.

\section{The Lambda Abstraction Strategy} 
\label{sec:LA}
While studying the tupling strategy and analyzing its power, a sentence by John 
Darlington, with whom I shared the office in Edinburgh, came often to my mind:
``After unfolding, having done some local improvements (such as the ones obtained
by the \sm{where} abstraction as shown in Section~\ref{sec:ProTrans} for the \sm{fusc} function), 
you need to fold.''
This {\it need for folding}~\cite{Dar81} is an important requirement. Folding steps
make the local improvements to be become global, so that they can be replicated at each level 
of recursion and thus become significant.

However, folding steps need matchings between expressions and these 
matchings may be sometimes impossible. Generalization of
constants to variables may allow matchings in some cases, 
but not always. In particular, when an expression 
should match one of its subexpressions, generalization of constant to variables
does not help. 
In those cases we have suggested to construct functions from
expressions~\cite{PeS89}. 
This is done by replacing the expression \sm{E[e]} where the 
subexpression~\sm{e} occurs, by the application \sm{(\lambda x. E[x])\, e}. 
We call this technique {\it lambda abstraction strategy} (or, as in other papers,
{\it higher-order abstraction}).

Let us see how lambda abstraction works in the following two examples taken 
from~\cite{PeS89}. The first 
example
refers to the 
following program {\it Reverse} for reversing a list, where \sm{[\,]}, 
\sm{:}, and \sm{@} denote the 
empty list, {\it cons}, and 
 {\it append} on lists, respectively. 
 
 \vspace{1mm}
{\small{
\noindent
\makebox[40mm][l]{${\mathtt{~~~1.~~rev([\,]) = [\,]}}$}  

\noindent
 ${\mathtt{~~~2.~~rev(a\!:\!\ell) = rev(\ell)~ @ ~[a]}}$ 

\noindent
\makebox[40mm][l]{${\mathtt{~~~3.~~[\,]~@~y = y}}$} 

\noindent
${\mathtt{~~~4.~~(a\!:\!\ell)~@~y = a : (\ell ~@~y)}}$
}}
\vspace{1mm}

%

\noindent
We want to derive a tail recursive definition of \sm{rev}.
 We need \sm{rev} to be 
the top operator of the right hand side of Eq.~\sm{2}, that is, 
\sm{rev(\ell)\,@\,[a]}, 
 and by induction we need that right hand side to be  
\sm{rev(\ell)}. There is a subexpression mismatch between \sm{rev(\ell)\,@\,[a]}
and \sm{rev(\ell)}. Then we proceed as follows: (i)~instead of \sm{rev(\ell)}, we
 consider \sm{rev(\ell)\,@\,[\,]}, (ii)~we 
generalize
the constant~\sm{[\,]} to the variable \sm{x}, thereby deriving 
\sm{rev(\ell)\,@\,x}, 
and (iii)~we abstract \sm{rev(\ell)\,@\,x} with respect to~\sm{x}, thereby 
deriving the function 
\sm{\lambda x.~rev(\ell)\,@\,x}.

The definition of the new function \sm{f(\ell)=_{def} \lambda x.~rev(\ell)\,@\,x}
is as follows.

\vspace{1mm}
\noindent
~~~\sm{5.~~f([\,]) = \lambda x.~rev([\,])\,@\,x =} \{by Eq.~\sm{1}\} = 
\sm{\lambda x.~[\,]\,@\,x = } 
\{by Eq.~\sm{3}\} = \sm{\lambda x.x } 
 
\noindent
~~~\sm{6.~~f(a\!:\!\ell) = \lambda x.\,rev(a\!:\!\ell) \,@\,x = \lambda x.\,(rev(\ell)\,@\,[a]) \,@\,x =} \{by associativity of 
\sm{@}\} = 

\hspace{13mm}= \sm{\lambda x.\,rev(\ell)\,@\,([a] \,@\,x) = } \sm{\lambda x.\,rev(\ell)\,@\,(a\!:\!x) = }
 \{by folding\} = \sm{\lambda x.\,(f(\ell)\,(a\!:\!x))} 

\vspace{1mm}
\noindent
We also have:

\vspace{1mm}
\noindent
\sm{~~~7.~~rev(\ell) = f(\ell)\,[\,]}
 
\vspace{1mm}
\noindent
The derived program (Eqs.~\sm{5}--\sm{7}) is more efficient than program 
(Eqs.~\sm{1}--\sm{4}) because
the expensive operation {\it append\/}  has been replaced by the cheaper
operation {\it cons}. Eqs.~\sm{5}--\sm{7} are basically equivalent to the 
program proposed in~\cite{Hug86} where a new representation for list has to 
be invented.

Note that the mechanization of the transformation we have now presented requires 
the use of associativity property for the append function. Thus, in general, 
it is important to have knowledge of the 
algebraic properties of the operations in use.

\vspace{1mm}
A second example refers to a problem proposed by Richard Bird~\cite{Bir84a}.
 Given a
binary tree \sm{t} we want to construct an isomorphic binary 
tree~\sm{\widetilde t} such that: (i)~\sm{t} and \sm{\widetilde t}
have the same multiset of leaves,  and (ii)~the leaves of 
\sm{\widetilde t},  when read from left to right, are in ascending order. 
One should derive a program which
construct~\sm{\widetilde t} by making one traversal only of the tree \sm{t}. 

In order to solve this program Richard Bird uses the so called {\it locally 
recursive programs}
whose semantics is quite complex and it is based on the 
{\it call-by-need} mode of evaluation.
By using the tupling and lambda abstraction strategies we will get the 
desired program
with the following advantages over Bird's solution: (i)~the use of 
{\it call-by-value} semantics, 
(ii)~the absence of local recursion, (iii)~the leaves are sorted {\it on the fly}, and (iv)~the computation of
components of tuples is done only when they are required for later computations.

By \sm{tip(n)} we denote a binary tree whose single 
leaf is the integer \sm{n}, 
and by \sm{t1\! \wedge\! t2} we denote a binary tree with children \sm{t1} and \sm{t2}.
By \sm{hd} and \sm{tl}
we denote, as usual, the {\it head\/} and {\it tail\/} functions on lists.
Our initial program is as follows. 

\vspace{1mm}
\noindent 
~~~\sm{1.~~TreeSort(t) = replace(t, sort(leaves(t)))}  where:

\vspace{1mm}
\noindent 
(i)~\sm{leaves(t)} returns the list of the leaves of the tree \sm{t},
(ii)~\sm{sort(\ell)} rearranges the list \sm{\ell} in ascending order from left to right, 
and
(iii)~\sm{replace(t,\ell)} uses in the left-to-right order the elements of the list
 \sm{\ell} to replace from left-to-right the leaves of the tree \sm{t}.
  
We assume that the length of \sm{\ell} is at least the number
of leaves in~\sm{t}. For instance, we have: 
\sm{TreeSort((tip(1)\!\wedge\! tip(2))\wedge tip(1)) \!=} 
\sm{(tip(1)\!\wedge\! tip(1))\wedge tip(2)}.
Here is the definition of the various functions required:

\vspace{2mm}
\noindent 
~~~\sm{2.~~leaves(tip(n)) = [n]}\nopagebreak  

\noindent 
~~~\sm{3.~~leaves(t1 \wedge t2) = leaves(t1) ~@~ leaves(t2)}  

\noindent 
~~~\sm{4.~~replace(tip(n), \ell) = tip(hd(\ell))}

\noindent 
~~~\sm{5.~~replace(t1 \wedge t2, \ell) = replace(t1, take(k, \ell)) \wedge replace(t2, 
drop(k, \ell))}
\hspace{5mm}\sm{where~ k\! =\! size(t1)}

\noindent 
~~~\sm{6.~~take(n,\ell) = if~ n= 0~ then~ [\,] ~else~take(n\!-\!1,\ell) ~@~  
[hd(drop(n\!-\! 1,\ell))]}

\noindent 
~~~\sm{7.~~drop(n,\ell) = if~  n = 0 ~then~ \ell ~else~  tl(drop(n \!-\!1, \ell))}

\vspace{2mm}
\noindent
For instance, \sm{take(2,[a,b,c,d,e])= [hd([a,b,c,d,e]), hd([b,c,d,e])]= [a,b]}
and 

\sm{drop(2,[a,b,c,d,e])=tl(tl([a,b,c,d,e]))= [c,d,e]}.

\vspace{1mm}
\noindent
As usual, given a list \sm{\ell}, we denote by \sm{length(\ell)} the number of elements in \sm{\ell}.
We assume that  \sm{0 \!\leq\! k \!\leq\! length(\ell)} holds when evaluating 
\sm{take(k, \ell)} and \sm{drop(k, \ell)}. 
For all list \sm{\ell}, for all \sm{0\!\leq\! n\!\leq\! length(\ell)}, we have \sm{\ell = take(n,\ell)\, @\, drop(n,\ell)}.  The function \sm{size(t)} returns the number of leaves in the tree \sm{t}. We have:

\vspace{1mm}
\noindent 
~~~\sm{8.~~size(tip(n)) = 1}

\noindent 
~~~\sm{9.~~size(t1  \wedge t2) = size(t1) + size(t2).}

\vspace{1mm}
\noindent
Here is the definition of \sm{sort} 
using \sm{merge} of two ordered lists:

\vspace{1mm}
\noindent
~~\sm{10.~~sort(\ell) = if~ \ell = [\, ] ~~then~~ [\,] ~~else~~ merge([hd(\ell)], 
sort(tl(\ell))).}

\noindent
~~\sm{11.~~merge([\,],\ell) = \ell}

\noindent
~~\sm{12.~~merge(\ell, [\,]) = \ell}

\noindent
~~\sm{13.~~merge(a\!:\!\ell1,\ b\!:\!\ell2) = if~~ a\!\leq\! b ~~then~~ a: 
merge(\ell1,\ b\!:\!\ell2) ~~else~~ b:merge(a\!:\!\ell1,\ \ell2)}

\vspace{1mm}
\noindent
Unfortunately, \sm{TreeSort(t)}  traverses the tree \sm{t} twice: a first 
visit is for 
collecting the leaves, and a second visit is for replacing them in ascending
 order. 

Now, let us start off the derivation of the one traversal algorithm by 
getting the 
inductive definition of \sm{TreeSort(t)}. From Eq.~\sm{1} we get:

\vspace{1mm}
\noindent
~~\sm{14.~~replace(tip(n), sort(leaves(tip(n)))) = replace(tip(n), sort([n])) = tip(n)}

\noindent
~~\sm{15.~~replace(t1\!\wedge\! t2, sort(leaves(t1\wedge t2))) = replace(t1, take(size(t1), 
\ell))~\wedge}

\hspace*{20mm}\sm{\wedge~replace(t2, drop(size(t1), \ell))  ~~~where~\ell = 
sort(leaves(t1\!\wedge\! t2))
}

\vspace{1mm}
\noindent
Now no folding step can be performed, because in 
\sm{replace(t1, take(size(t1), \ell))} 
the subexpression \sm{take(size(t1), \ell)} does not match 
\sm{sort(leaves(t1))}. 
Similarly, for the subtree \sm{t2}, instead of \sm{t1}. By the lambda 
abstraction we generalize the 
mismatching subexpression to the list variable \sm{z}, and we introduce the 
function 
\sm{\lambda z.~replace(t,z)} whose definition is as follows (the details are 
in~\cite{PeS89}):

\vspace{1mm}
\noindent
~~\sm{16.~~\lambda z.~replace(tip(n), z) =  \lambda z.~tip(hd(z))}\nopagebreak

\noindent
~~\sm{17.~~\lambda z.~replace(t1\!\wedge\! t2,z) = 
\lambda z.\,((\lambda y.~replace(t1, y) 
~take(k, z)) \ \wedge}\nopagebreak

~~\hspace*{15mm}\sm{ \wedge~((\lambda y.~replace(t2,y)) ~ 
drop(k,z))) ~~~where~~ k = size(t1)}

\vspace{1mm}
\noindent
The functions \sm{\lambda z.~replace(t,z)} and 
\sm{sort(leaves(t))} visit the same tree~\sm{t}. We apply the tupling 
strategy and we define the function:

\sm{T(t) =_{def}\langle \lambda z.~replace(t, z),\ sort(leaves(t))\rangle}

\noindent
whose explicit definition is:

\vspace{1mm}
\noindent
~~\sm{18.~~T(tip(n)) = \langle \lambda z.~tip(hd(z)), ~[n]\rangle}

\noindent
~~\sm{19.~~T(t1\! \wedge\! t2)\! =\! \langle \lambda z.\,((a1~take(size(t1), z)) 
\wedge (a2~drop(size(t1), z))), ~merge(b1,b2)\rangle}\nopagebreak

\hspace{30mm}\sm{where~~\langle a1,b1\rangle \!=\! T(t1) ~~and~~ 
\langle a2,b2\rangle \!=\! T(t2)}

\vspace{1mm}
\noindent
Now \sm{T(t1)}, \sm{take(size(t1), z)}, and \sm{drop(size(t1), z)}
visit the same tree~\sm{t1}. We apply the tupling strategy and we introduce 
the new function:

\vspace*{1mm}
\sm{U(t,y) =_{def} \langle \lambda z.~replace(t,z), ~sort(leaves(t)), 
~take(size(t), y), ~drop(size(t), y) \rangle}

\vspace*{1mm}
\noindent
We get the following explicit definition for \sm{U(tip(n),y)}:

\vspace{1mm}
\sm{U(tip(n),y) =\langle \lambda z.~tip(hd(z)), ~[n], ~[hd(y)], ~tl(y)\rangle}

\vspace{1mm}
\noindent
However, when looking for the explicit definition of 
\sm{U(t1\!\wedge\! t2 ,y)} we get again
a subexpression mismatch (see~\cite{PeS89}) and we use again 
lambda abstraction for the
last two components of the 4-tuple \sm{U(t,y)}. Thus, we introduce the following function:

\vspace*{1mm}
\sm{V(t) =_{def}\!\langle \lambda z.replace(t,z),\ sort(leaves(t)),\  
\lambda z.take(size(t),z),\ \lambda z.drop(size(t),z) \rangle}

\vspace*{1mm}
\noindent
whose explicit definition is:

\vspace{1mm}
\noindent
~~\sm{20.~~V(tip(n)) =\langle \lambda z.~tip(hd(z)), ~[n], ~\lambda z.~[hd(z)], 
~\lambda z.~tl(z)\rangle}

\noindent
~~\sm{21.~~V(t1 \!\wedge\!t2) \!=\! \langle  \lambda z.((a1\,(c1\,z)) \wedge 
(a2~(d1~z))), ~merge(b1,b2),
~\lambda z.((c1~z)\, @\, (c2\,(d1\,z)\!)\!),~ \lambda z.(d2\,(d1\,z))  \rangle }

\hspace*{25mm}\sm{where~\langle a1,b1,c1,d1\rangle \!=\! V(t1) ~~and~~
\langle a2,b2,c2,d2\rangle \!=\! V(t2)}


\vspace{1mm}
\noindent
We get the following program such that
for all trees \sm{t}, \sm{NewTreeSort(t) = TreeSort(t)} (see Eq.~\sm{1}):

\vspace{1mm}
\noindent
~~\sm{22.~~NewTreeSort(t) = (a2~b2) ~~~~~where~\langle a2, b2\rangle = T(t)}\nopagebreak

\noindent
~~\sm{18.~~T(tip(n)) = \langle \lambda z.~tip(hd(z)), ~[n]\rangle}

\noindent
~~\sm{23.~~T(t1\! \wedge\! t2)  =  \langle \lambda z.\,((a1~c1, z)) \wedge (a2~d1, z))), ~merge(b1,b2)\rangle} 

\hspace{25mm}\sm{where~~\langle a1,b1,c1,d1\rangle \!=\! V(t1) ~~and~~ \langle a2,b2\rangle \!=\! T(t2)}

\vspace{1mm}
\noindent
together with Eqs.~\sm{20} and~\sm{21} for the function \sm{V(t)}.

A further improvement of this program can be made by avoiding the append function \sm{@} occurring in Eq.~\sm{21}. One can use the same technique of lambda abstraction 
shown in the {\it Reverse} example at the beginning of this section. 
We consider a
variant of the function \sm{V(t)} whose 3rd component is the abstraction \sm{\lambda 
z\,x.~take(size(t),z)\,@\,x}, instead of \sm{\lambda z.~take(size(t),z)}. 
The function \sm{T^{*}(t)} is like~\sm{T(t)}, but uses \sm{V^{*}(t)}, instead of 
\sm{V(t)}. 
We get the following final program such that
for all trees \sm{t}, \sm{NewTreeSort^{*}(t) = TreeSort(t)}:

\vspace{2mm}
\noindent
~~\sm{22^{*}\!.~~NewTreeSort^{*}(t) = (a2~b2) ~~~~~where~\langle a2, 
b2\rangle = T^{*}(t)}

\noindent
~~\sm{18^{*}\!.~~T^{*}(tip(n)) = \langle \lambda z.~tip(hd(z)), ~[n]\rangle}

\noindent
~~\sm{23^{*}.~~T^{*}(t1\! \wedge\! t2) = \langle \lambda z.\,((a1~(c1(z,
[\,]))) \wedge (a2~(d1~z))), ~merge(b1,b2)\rangle}\nopagebreak

\hspace{32mm}\sm{where~~\langle a1,b1,c1,d1\rangle \!=\! V^{*}(t1) ~~and~~ 
\langle a2,b2\rangle \!=\! T^{*}(t2)} 

\noindent
~~\sm{20^{*}.~~V^{*}(tip(n)) =\langle \lambda z.~tip(hd(z)), ~[n], ~\lambda 
z\,x.~hd(z)\!:\!x, 
~\lambda z.~tl(z)\rangle}

\noindent
~~\sm{21^{*}.~~V^{*}(t1 \!\wedge\!t2) = \langle  \lambda z.\,((a1~(c1(z,
[\,]))) \wedge (a2~(d1~z))), ~~merge(b1,b2),} \nopagebreak

~~\hspace*{23.5mm}\sm{~\lambda z\,x.\,(c1(z,c2((d1~z),x))),} ~\sm{ \lambda z.
(d2~(d1~z))  \rangle}\nopagebreak

\hspace*{32mm}\sm{where~ \langle a1,b1,c1,d1\rangle \!=\! V^{*}(t1) ~~and~~ 
\langle a2,b2,c2,d2\rangle \!=\! V^{*}(t2)}

\vspace{2mm}
\noindent
Computer experiments performed at the time of writing the paper~\cite{PeS89} 
from which 
we take this example, show that the computation of
the final function \sm{NewTreeSort^{*}(t)} is faster than the one of the 
initial function 
\sm{TreeSort(t)} for trees whose size is greater than 
about 30. For trees of smaller size the overhead of dealing with functions is not
compensated by the fact that the input tree is visited once only.

Note also that since lambda expressions do not have free variables,
we can operate on them by using pairs of bound variables and function 
bodies, instead of the more expensive closures. 
Thus, for instance, \sm{\lambda z.\,expr}
can be represented by the pair~\sm{\langle z, expr\rangle}.

\smallskip
Some years later Maurizio Proietti and I have studied the application of the
lambda abstraction strategy in the area of logic programming.
As in functional programs where we have lambda expressions denoting 
functions,
in logic programming we should have terms denoting goals, and thus 
goals should be allowed to occur as arguments of predicates. To allow 
goals as arguments,
we have proposed a novel logic language, we have defined its semantics, 
and we have provided for it a set of 
unfold/fold transformations rules, together with some goal replacement rules, 
such as the one stating the equivalence 
of the goal $g \wedge \mathit{true}$ with the goal $g$~\cite{PeP99b,PeP04}. 
Those rules have been proved correct.

Here is an example of
efficiency improvement obtained by program transformation in this novel 
language. This transformation has not been mechanized, but we believe that it 
is not hard to do it.  Details can be found in~\cite[Section~7.1]{PeP04}. 
Let us consider
a program which given a binary tree (either $l(N)$ or $t(L,N,R)$), (i)~flips all its left and right subtrees, and 
(ii)~checks in a subsequent traversal of the tree, whether or not all labels are 
natural numbers.


\vspace{1mm}
\Feq{1.}\mathit{flipcheck}(X,Y)\leftarrow \mathit{flip}(X,Y),\
\mathit{check}(Y) \Nex{2.}\mathit{flip}(l(N),l(N))\leftarrow
\Nex{3.}\mathit{flip}(t(L,N,R),t(\mathit{FR},N,\mathit{FL}))\leftarrow 
\mathit{flip}(L,\mathit{FL}),\
        \mathit{flip}(R,\mathit{FR})
\Nex{4.}\mathit{check}(l(N))\leftarrow \mathit{nat}(N)
\Nex{5.}\mathit{check}(t(L,N,R))\leftarrow \mathit{nat}(N),\
        \mathit{check}(L),\ \mathit{check}(R)
\Nex{6.}\mathit{nat}(0)\leftarrow
\Nex{7.}\mathit{nat}(s(N))\leftarrow \mathit{nat}(N)
\eeq

\vspace{1mm}

\noindent
We derived the following program which traverses the input tree only
once and uses the continuation passing style: 

\vspace{1mm}
\Feq{8}\mathit{flipcheck}(X,Y)\leftarrow
\mathit{newp}(X,Y,G,\mathit{true},G)

\Nex{9}\mathit{newp}(l(N),l(N),G,C,D) \leftarrow
         \mathit{eq}\U\mathit{c}(G,\mathit{nat}\U\mathit{c}(N,C),D)

\Nex{10}\mathit{newp}(t(L,N,R),t(\mathit{FR},N,\mathit{FL}),G,C,D) \leftarrow
\Nex{}\hspace*{20mm}\mathit{newp}(L,\mathit{FL},U,C,\ 
                    \mathit{newp}(R,\mathit{FR},V,U,\ 
             \mathit{eq}\U\mathit{c}(G,\mathit{nat}\U\mathit{c}(N,V),D)))
\Nex{11}\mathit{nat}\U\mathit{c}(0,C)\leftarrow C
\Nex{12}\mathit{nat}\U\mathit{c}(s(N),C)\leftarrow
         \mathit{nat}\U\mathit{c}(N,C)
\eeq

\vspace{1mm}
\noindent
For the predicate $\mathit{eq}\U\mathit{c}$ we assume that: 
$\vdash \forall\,( {\it eq}\U{c}(X,Y,C)\, \leftrightarrow\, ((X\!=\!Y) \wedge C)))$. 

\vspace*{-2mm}

\section{Communications and Parallelism} 
\label{sec:CommPar}
While at Edinburgh I had the privilege of attending a course on the 
Calculus of
Communicating Systems (CCS) by Professor Robin Milner 
(1934-2010)~\cite{Mil89}. 
I remember the day when Robin Milner and  Gordon Plotkin
 decided the name to be given
to this new calculus. As I was told, they first  decided that the name should 
have been of three letters only! I appreciated the beauty of the calculus
which resembles a development of lambda calculus. 
The application of {\it a 
function} \sm{\lambda x.\,e[x]} to {\it an argument} \sm{a} can, indeed, 
be understood as  a
communication which takes place between: (i)~the `function agent' and 
(ii)~the `argument agent' through the `port' named \sm{\lambda}. After their 
communication, which is called a {\it
handshaking}, the agents continue their respective activities, namely, 
(i)~the function agent does the evaluation of \sm{e[a]}, that is, the body 
\sm{e[x]} of the function 
where the variable \sm{x} have been bound to the value \sm{a}, 
and (ii)~the argument agent does nothing, that is, it become 
the {\it null-agent} (indeed,
for the rest of  the computation, the argument has nothing left to do).

At about the same time, Professor Tony
Hoare in Oxford was developing his calculus of Concurrent Sequential 
Programs 
(CSP)~\cite{Hoa78}. I remember a visit that Tony Hoare made to Robin
Milner at Edinburgh and 
the stimulating seminar Hoare gave on CSP on that occasion. 

In subsequent years, I thought of exploring the power of communications and
parallelism in functional programming, also because the various components of the 
tuples introduced by
the tupling strategy can be computed in parallel. These components
can be considered as independent agents which may synchronize at the
end of their computations. 
During those years, the notion of communicating agents was emerging quite 
significantly in various programming paradigms.

Andrzej Skowron and I did some work in this area and 
we proposed (some variants of) a 
functional language with communications~\cite{PeS85a,PeS85b}.
Each function call is assumed to be an {\it agent}, that is, a triple of the
form \sm{\langle x,m\rangle\!::\!expr}, where \sm{x} is its name, 
\sm{m} is its {\it message}, that is, its local information, and  \sm{expr} is its expression, that is, the
task it has to perform. The operational semantics of the language 
is based on the conditional rewriting of sets (or multisets) of agents,
similarly to what is done in coordination 
languages (see, for instance,~\cite{GeC92}).

As an example of a functional program with communications which we 
proposed,
let us consider the following program for computing the familiar 
Fibonacci function. 

The variable~\sm{x} ranges over agent names which are strings
constructed from~\sm{x} as the following grammar indicates:
\sm{x ~::=~ \varepsilon ~\mid~ x.0 ~\mid~ x.1}.
The left and right son-calls of the agent whose name is~\sm{x} have names 
\sm{x.0}
and \sm{x.1}, respectively. By default, the name of the agent of the 
initial
function call is the empty string~\sm{\varepsilon}. 

In our example, the 
variables 
\sm{ms} and \sm{ms1} range over 
the three message constants: \sm{R}~(for {\it ready}),
\sm{R1}~(for {\it{ready\/}\sm{1}}), and~\sm{W}~(for {\it {\it{wait}}}).
Agents with messages \sm{R} and \sm{R1} may make rewritings, while 
agents with message \sm{W} cannot (see Rules~\sm{1}--\sm{4} below).
The variables \sm{n}  and \sm{val} range over integers and the variable 
\sm{exp} ranges over integer expressions.

\vspace{2mm}
\noindent
~~\sm{1.~~\big\{\langle x,ms\rangle\!::\!fib(0)\big\}  ~\Rightarrow~ 
\big\{\langle x,ms\rangle\!::\!0\big\}}    \hspace{18mm} 
\sm{if~ ~ms\!=\!R ~~or~~ ms\!=\!R1}\vspace{1mm}

\noindent
~~\sm{2.~~\big\{\langle x,ms\rangle\!::\!fib(1)\big\}  ~\Rightarrow~ 
\big\{\langle x,ms\rangle\!::\!1\big\}}   \hspace{18mm} 
\sm{if~ ~ms\!=\!R ~~or~~ ms\!=\!R1}\vspace{1mm}

\noindent
~~\sm{3.~~\big\{\langle x,R\rangle\!::\!fib(n\!+\!2)\big\}  ~\Rightarrow~ 
\big\{\langle x,R\rangle\!::\!+(x.0,x.1),\ \langle x.0,R\rangle\!::\!fib(n\!+\!1),
\ \langle x.1,R1\rangle\!::\!fib(n)\big\}}  \hfill \sm{if ~n\!\geq\! 
0}~~~\vspace{1mm}


\noindent
~~\sm{4.~~\big\{\langle x,R1\rangle\!::\!fib(n\!+\!2)\big\}  ~\Rightarrow~ 
\big\{\langle x,R1\rangle\!::\!+(x.0,x.1),\ \langle x.0,W\rangle\!::\!fib(n\!+\!
1),\ \langle x.1,R\rangle\!::\!fib(n)\big\}}   \hfill \sm{if ~n\!\geq\! 
0}~~~\vspace{1mm}


\noindent
~~\sm{5.~~\big\{\langle x.0,ms\rangle\!::\!val,\ \langle x,ms1\rangle\!::+
(x.0,exp)\big\}
~\Rightarrow~ 
\big\{\langle x,ms1\rangle\!::+(val,exp)\big\}}\vspace{1mm}

\noindent
~~\sm{6.~~\big\{\langle x.1,ms\rangle\!::\!val,\ \langle x,ms1\rangle\!::+
(exp,x.1)\big\}
~\Rightarrow~ 
\big\{\langle x.1,ms\rangle\!::\!val,\ \langle x,ms1\rangle\!::+(exp,val)
\big\}}\vspace{1mm}

\noindent
~~\sm{7.~~\big\{\langle x.0.1,R1\rangle\!::\!val,\ \langle 
x.1.0,W\rangle\!::exp\big\}
~\Rightarrow~ 
\big\{\langle x.0.1,R1\rangle\!::\!val,\ \langle x.1.0,R\rangle\!::val\big\}}\vspace{1mm}

\vspace{1mm}
\noindent
Rules~\sm{1} and~\sm{2} are the expected ones for computing
\sm{fib(0)} and \sm{fib(1)}. The recursive call of 
\sm{fib(n\!+\!2}) has two variants (see Rules~\sm{3} and \sm{4}) so to be able to evaluate 
the call of agent~\sm{x.0.1} in a different way than that of 
agent~\sm{x.1.0}. 
The expression \sm{+(x.0,x.1)} has the effect that, once the values of the son-calls
are evaluated and sent to the father-call, according to Rules~\sm{5} 
and~\sm{6},
then the father-call silently performs the sum of the values it has received. 
Rule~\sm{7} sends the value computed by agent \sm{x.0.1} to 
agent \sm{x.1.0}. This communication is correct and improves efficiency.
Indeed,  by our program the value of \sm{fib(n\!-\!1)} which is needed 
for computing \sm{fib(n\!+\!1)} and \sm{fib(n)}, is computed once only. 
Note, in fact, that one of the
two agents which have to compute \sm{fib(n\!-\!1)}, has the message~\sm{W} and
cannot make further rewritings.

We have considered the problem of how to modify the rules of the programs  
when acquiring knowledge of new facts about the functions to be evaluated
for improving program efficiency. In the case of the Fibonacci 
function, one such fact may be the equality of the expressions 
to be computed by the agents \sm{x.0.0} and \sm{x.1}.  

Note that the above Rules~\sm{1}--\sm{7} do not perform the 
on-the-fly garbage collection of the agents because right-sons are not erased.
To overcome this problem one may use more complex messages~\cite{PeS85b}
so that every
agent knows the agents which are waiting for receiving the value it computes. 
If there are 
none, the agent may be erased once it has sent its value to the father-call.

Note also that, if instead of Rule~\sm{6}, we use the simpler equation:

\vspace{.5mm}
\noindent
~~\sm{{\widetilde{6}}.~~\big\{\langle x.1,ms\rangle\!::\!val,\ 
\langle x,ms1\rangle\!::+(exp,x.1)\big\}
~\Rightarrow~ 
\big\{\langle x,ms1\rangle\!::+(exp,val)\big\}}

\vspace{.5mm}
\noindent
deadlock may be generated.  We have also proposed a modal logic for proving
correctness of our functional programs with agents and 
communications~\cite{PeS83} and, in particular, the absence of deadlock.
Unfortunately, no implementation of our language proposal and its
modal logic has been done.

\smallskip
Concerning a more theoretical study of parallelism and communications,
Anna Labella and~I considered categorical models for calculus with
handshaking communications both in the case of CCS~\cite{Mil89} and 
CSP~\cite{Hoa78}. We were inspired by the definition of
the cartesian closed categories 
for providing models of the lambda calculus. 

We followed an approach
different from Winskel's one~\cite{Win84}. We did not give an {\it a priori}
definition of a categorical structure, where the embedding of the algebraic 
models of CCS or CSP might not be completely satisfactory. 
We started, instead, from 
the algebraic models, based on labelled trees of various kinds, 
and we defined suitable
categories of labeled trees where one can interpret all the basic operations of 
CCS and CSP. In a sense, we followed the approach presented many years 
earlier by Rod Burstall for the description of flowchart programs~\cite{Bur72}. 
The details of our categorical constructions can be found 
in~\cite{Ka&90,LaP85}. 

In some models of ours we used 
{\it enriched categories}~\cite{Kel82}. 
An enriched category is a category where the sets of morphisms associated with
the pairs of objects, are replaced by objects from a fixed monoidal category.
For lack of space we will not enter into the details here.


\section{Transformation and Verification in Logic Programming} 
\label{sec:TransfVerifinLogic}
While studying at Edinburgh, I thought of applying the transformation 
methodology to CCS agents. 
I remember talking to Robin Milner about this idea. He did not show much interest 
maybe because for him it was more important to first acquire a good 
understanding the equivalences between terms in the 
CCS calculus, before applying them to the transformations of agents which, of 
course, should be equivalence preserving.

Then, I thought of applying program transformation to the area of logic programming
which I first studied during my Ph.D.~research at Edinburgh. 
At that time William Clocksin and Chris Mellish were writing their 
popular book on Prolog~\cite{ClM84}. 
I remember reading some parts of a draft of the book. 
Also I had the chance of looking at David Warren's report 
on how to compile logic programs~\cite{War77}. I also read his 
paper comparing the Prolog implementation with Lisp~\cite{Wa&77}
and the later report on the Warren Abstract Machine~\cite{War83}. 
From those days I still remember David's kindness, his cooperation with
Fernando Pereira, and his love for plants and flowers.

\smallskip
A few years later, when back in Italy, I was introduced by Anna Labella
to her former student Maurizio Proietti who, not long before, 
had graduated in Mathematics at Rome University \mbox{`La Sapienza'},
defending a thesis on Category Theory.
I spoke to Maurizio and I introduced him to logic programming~\cite{Llo87}.
I also encouraged him to work in the field of logic program transformation. 
He kindly accepted. The basis of his work was a paper by Hisao Tamaki 
and Taisuke Sato~\cite{TaS84}
that soon afterwards became the standard reference 
for logic program transformation.

That was the beginning of Maurizio's cooperation with me. He was first
funded by a research grant from the private company Enidata (Rome) and soon later, he became
a researcher of the Italian National Research Council in Rome.
We first considered some techniques for 
finding the {\it eureka predicates}, that is, the predicate definitions 
to be introduced during program transformation~\cite{PrP90b}.

Besides the definition introduction, unfolding, and folding rules, 
we have used for our transformations
a rule called {\it Generalization\,$+$\,Equality Introduction} (see 
also~\cite{BoM75} for
a similar rule when proving theorems in functional programs). By this 
rule, a clause of the form $H \If A_{1},\ldots,A_{n}$ is generalized to the 
clause
$H \If {\mathit{GenA}}_{1},\ldots,{\mathit{GenA}}_{n},$ $X_{1}\!=\!t_{1}$, 
$\ldots, X_{n}\!=\!t_{r}$, where 
$({\mathit{GenA}}_{1},\ldots$, ${\mathit{GenA}}_{n})\,
\vartheta\!=\!(A_{1},\ldots,A_{n})$ being
$\vartheta$ the substitution
$\{X_{1}/t_{1},\ldots,$ $ X_{r}/t_{r}\}$.

We have also introduced: (i)~the class of {\it non-ascending programs}, where, 
among other properties,
each variable should occur in an atom at most once, (ii)~the {\it synchronized 
descent rule}
(SDR) for driving the unfolding steps by selecting the
atoms to be unfolded, and (iii)~the {\it loop absorption strategy} for the 
synthesis 
of the eureka predicates. We have also
characterized classes of programs in which that strategy is guaranteed to be successful.

\smallskip
Let us see a simple example of application of the loop absorption strategy. 
Here is a program, called $\mathit{Comsub}$, for computing common subsequences of
 lists.
 

\vspace{1mm}
\Feq{1.}\mathit{comsub}(X,Y,Z)\leftarrow \mathit{sub}(X,Y),\ \mathit{sub}(X,Z)
\vspace{.5mm} 
\Nex{2.}\mathit{sub}([\,],X)\leftarrow\vspace{.5mm}
\Nex{3.}\mathit{sub}([A|X],[A|Y])\leftarrow \mathit{sub}(X,Y)\vspace{.5mm}
\Nex{4.}\mathit{sub}(X,[A|Y])\leftarrow \mathit{sub}(X,Y)\vspace{.5mm}
\eeq

\vspace{1mm}
\noindent
where $\mathit{sub}(X,Y)$  holds iff $X$ is a sublist of $Y$. The order of 
the elements should be preserved, but the elements in $X$ need not to be 
consecutive in $Y$. For instance, [1,2] is a sublist of [1,3,2,3], while [2,1] is 
not.
We want to derive a program where the double visit of the list $X$ in clause~1 is avoided.

First, we make the given program to be non-ascending by replacing clause 3 by 
the following clause: 

\vspace{1mm}
3.1~${\mathit{sub}([A|X],[A1|Y]) \!\leftarrow\! A\!=\!A1, 
\mathit{sub}(X,Y)}$

\begin{figure}[ht!]
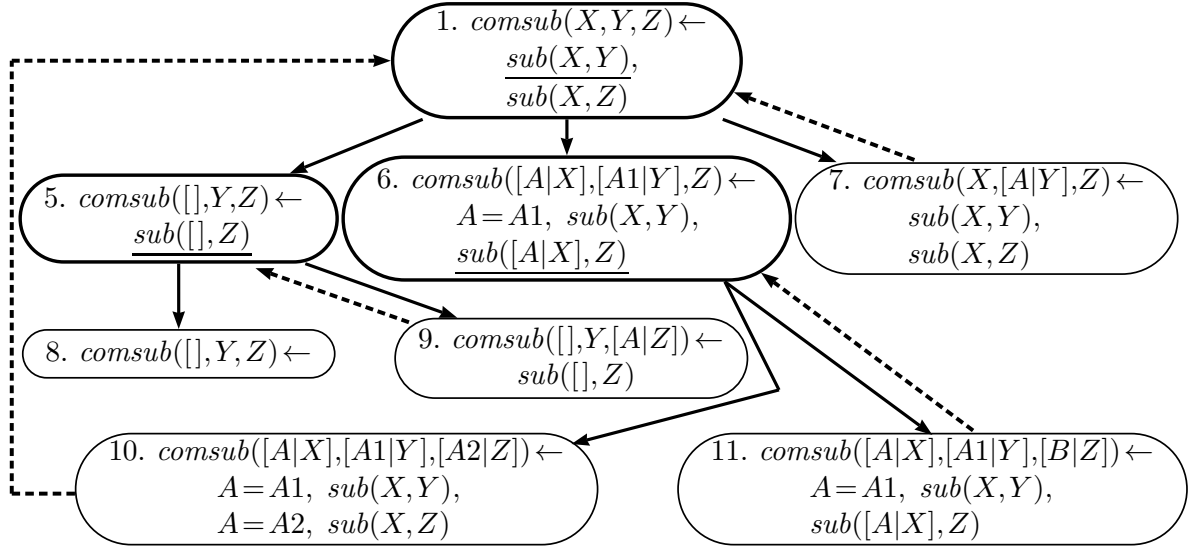

\begin{center}
\VCDraw{%
\begin{VCPicture}{(-6.7,-1)(20,11.5)} 
\SetEdgeArrowWidth{8pt}


\ChgStateLineWidth{1.2}
\FixStateDiameter{26mm}
\StateVar[\begin{array}{l} ~~1.~{\mathit{comsub}}(X,Y,Z)\! \leftarrow ~~\\ 
~~~~~~~~~\underline{{\mathit{sub}}(X,Y)},\\ 
~~~~~~~~~{\mathit{sub}}(X,Z) \end{array}]{(5.9,9.5)}{1}

\FixStateDiameter{20mm}
\StateVar[\begin{array}{l} 5.~{\mathit{comsub}}(\nil,\!Y,\!Z)\! \leftarrow ~~\\ 
~~~~~~~~~\underline{{\mathit{sub}}(\nil,Z)} \end{array}]{(-2.7,6)}{5}

\FixStateDiameter{28mm}
\StateVar[\begin{array}{l} ~6.~{\mathit{comsub}}(\AX,\!\AlY,\!Z)\! \leftarrow ~~\\ 
~~~~~~~~~A\!=\!A1,\ {{\mathit{sub}}(X,Y)},\\
~~~~~~~~~\underline{{\mathit{sub}}(\AX,Z)} \end{array}]{(5.9,6)}{6}

\SetStateLineStyle{solid}
\ChgStateLineWidth{.6}
\FixStateDiameter{11mm}
\StateVar[\begin{array}{l} 8.~{\mathit{comsub}}(\nil,Y,Z) \!\leftarrow 
\end{array}]{(-2.7,3)}{8}

\FixStateDiameter{18mm}
\StateVar[\begin{array}{l} 9.~{\mathit{comsub}}(\nil,\!Y,\!\AZ)\! \leftarrow\\
~~~~~~~~~~{{\mathit{sub}}(\nil,Z)}  \end{array}]{(6.0,2.9)}{9}

\FixStateDiameter{25mm}
\StateVar[\begin{array}{l} ~7.~{\mathit{comsub}}(X,\!\AY,\!Z) \!\leftarrow ~~\\ 
~~~~~~~~~{{\mathit{sub}}(X,Y)},\\
~~~~~~~~~{{\mathit{sub}}(X,Z)} \end{array}]{(15.2,6)}{7}

\FixStateDiameter{25mm}
\StateVar[\begin{array}{l} ~10.~{\mathit{comsub}}(\AX,\!\AlY,\!\AsZ)\! \leftarrow ~~\\ 
~~~~~~~~~~~~A\!=\!A1, \ {{\mathit{sub}}(X,Y)},\\
~~~~~~~~~~~~A\!=\!A2, \  {{\mathit{sub}}(X,Z)} \end{array}]{(0.8,0)}{10}

\FixStateDiameter{25mm}
\StateVar[\begin{array}{l} ~11.~{\mathit{comsub}}(\AX,\!\AlY,\!\BZ) \!\leftarrow 
~~\\ 
~~~~~~~~~~~A\!=\!A1, \ {{\mathit{sub}}(X,Y)},\\
~~~~~~~~~~~{{\mathit{sub}}(\AX,Z)} \end{array}]{(14,0)}{11}

\ChgEdgeLineWidth{2}
\EdgeL{1}{5}{} \EdgeL{1}{6}{}  \EdgeL{1}{7}{} 
\EdgeL{5}{8}{} \EdgeL{5}{9}{}   

\Point{(9.4,4.6)}{6c}       
\Point{(10.6,2.2)}{10c}
\Point{(6,1)}{10cc}
\SetEdgeArrowWidth{0pt}
\EdgeL{6c}{10c}{} 
\SetEdgeArrowWidth{8pt}
\EdgeL{10c}{10cc}{}

\Point{(9.4,4.6)}{6b}       
\Point{(14,1.2)}{11b}
\EdgeL{6b}{11b}{} 

\ChgEdgeLineStyle{dashed}    
\ChgEdgeLineWidth{2.5}
\ChgEdgeArrowWidth{8pt}

\Point{(2.0,9.5)}{1b0}     
\Point{(-6.4,9.5)}{1b1}
\Point{(-6.4,-.1)}{1b2}
\Point{(-5,-.1)}{1b3}
\SetEdgeArrowWidth{0pt}
\EdgeL{1b3}{1b2}{}  \EdgeL{1b2}{1b1}{}  
\ChgEdgeArrowWidth{8pt}
\EdgeL{1b1}{1b0}{}   

\Point{(-1.,4.9)}{5a}    
\Point{(2.4,3.7)}{9a}
\EdgeL{9a}{5a}{} 

\Point{(10.2,4.8)}{6a}    
\Point{(14.9,1.3)}{11a}
\EdgeL{11a}{6a}{} 

\Point{(9.6,8.8)}{1a}  
\Point{(13.6,7.3)}{7a}
\EdgeL{7a}{1a}{}

\end{VCPicture}
}
\end{center} \normalsize \caption{An upper portion of the unfolding tree for 
{\it Comsub}$1$.\label{fig:CommSubSeqTree}}
\end{figure}

\vspace{1mm}
\noindent
Let {\it Comsub}$1$ be the set $\{1,2,3.1,4\}$ of clauses.
In Figure~\ref{fig:CommSubSeqTree} we have depicted an
upper portion of the unfolding tree for {\it Comsub}$1$.
In that figure we have underlined the atoms which are 
unfolded. 
Solid down-arrows denote unfolding, and dashed up-arrows denote loops which 
suggest the 
definition clauses needed for folding, as we will explain. In clause~6 we 
unfold the 
atom~${\mathit{sub}}(\AX,Z)$ which is selected by the SDR rule. 
Indeed, by the synchronized descent rule, in clause~6 we have to unfold that atom, because in its 
ancestor-clause~1 we have 
unfolded the other atom~${\mathit{sub}}(X,Y)$ occurring in the body of that ancestor.
Unfolding is stopped when the recursive 
defined atoms in the body of a leaf-clause, say~$L$, are subsumed by 
the body of an ancestor-clause, say~$A$. 
In this case we say that a loop of the form $\langle A,L\rangle$ 
has been detected. 
Details can be found in~\cite{PrP90b}. 

According to the loop absorption strategy, for each detected loop $\langle A,L\rangle$  
we introduce a new definition clause~$D$ 
so that the bodies of both 
clauses~$A$ and~$L$ can be folded using $D$.
The loops $\langle 1,10\rangle$ and $\langle 1,7\rangle$ need not a new 
definition because we have clause~1 defining {\it comsub}.
The loops $\langle 5,9\rangle$ and $\langle 6,11\rangle$ require the following 
two new predicate definitions

\vspace{1mm}
\makebox[90mm][l]{$\mathit{newsub}(Z)\!\leftarrow\! \mathit{sub}(\nil,\!Z)$} for loop $\langle 5,9\rangle$ \nopagebreak

\vspace{.5mm}
\makebox[90mm][l]{$\mathit{newcomsub}(A,\!X,\!Y,\!Z)\!\leftarrow\! \mathit{sub}(X,\!Y),\  
\mathit{sub}([A|X],\!Z)$} for loop $\langle 6,11\rangle$


\vspace{1mm}
\noindent
By performing the unfolding and folding steps which correspond to the subtrees
rooted in clauses~1, 5, and 6 of Figure~\ref{fig:CommSubSeqTree}, we get the 
explicit definitions of the predicates $\mathit{newsub}$ and
$\mathit{newcomsub}$. 

Eventually, by simplifying the equalities, we get the 
following program:

\vspace*{1mm}

\Feq{5.}\makebox[90mm][l]{$\mathit{comsub}([\,],Y,Z)\leftarrow$} (*)\nopagebreak
\Nex{6.}\mathit{comsub}([\,],Y,[A|Z])\leftarrow \mathit{newsub}(Z)
\Nex{7.}\makebox[90mm][l]{$\mathit{comsub}([A|X] ,[A|Y],Z)\leftarrow \mathit{newcomsub}(A,X,Y,Z)$} (*)
\Nex{8.}\makebox[90mm][l]{$\mathit{comsub}(X,[A|Y],Z) \leftarrow \mathit{comsub}(X,Y,Z)$} (*)
\Nex{9.}\makebox[90mm][l]{$\mathit{newcomsub}(A,X,Y,[A|Z]) \leftarrow  \mathit{comsub}(X,Y,Z)$} (*)
\Nex{10.}\makebox[90mm][l]{$\mathit{newcomsub}(A,X,Y,[B|Z]) \leftarrow 
        \mathit{newcomsub}(A,X,Y,Z)$} (*)
\Nex{11.}\mathit{newsub}(Z)
\Nex{12.}\mathit{newsub}([A|Z]) \leftarrow \mathit{newsub}(Z)
\eeq

\vspace{1mm}
\noindent
Now clause~6 is subsumed by clause~5 and can be erased. Then, also clauses~11 and
12 can be erased and the final program is made out of the marked clauses~5, 7--10 
only. This final program is equal to the one derived by Tamaki-Sato~\cite{TaS84}. 
Note that our derivation does not rely
on human intuition and can easily be mechanized. 
The computation of all solutions of the goal $\mathit{comsub}(X,Y,Z)$,
where~$X$ is a free variable and $Y$ and $Z$ are ground lists of 10 elements, is
about~6~times faster when using the final program, instead of the initial 
one~\cite{PrP90b}. A development of the technique we have now illustrated 
can be found in~\cite{PrP91b}. 

\smallskip
The following example, taken from a paper of ours~\cite{PeP02a} written some years
later in honor of Professor Robert Kowalski, 
shows an application of the program transformation methodology also to the case
when clauses may have negated atoms in their body. 
For that kind of logic programs, called  {\it locally stratified logic programs},
we have also provided the transformation rules that can be applied 
and we have shown that they are correct, in the sense that they preserve the
perfect model semantics. 
The details on the rules and the definition of the perfect model semantics
can be found in~\cite{PeP02a}.

Let us consider the following program~\emph{CFParser} for deriving a word 
generated by
a given context-free grammar over the alphabet $\{a,b\}$:

\vspace{2mm}
\Feq{1.}\mathit{derive}([\,],[\,])\leftarrow \\[.5mm]
\Nex{2.}\mathit{derive}([A|S],[A|W])\leftarrow \mathit{terminal}(A),\, 
\mathit{derive}(S,W)\\[.5mm]
\Nex{3.}\mathit{derive}([A|S],W)\leftarrow \mathit{nonterminal}(A),\  
\mathit{production}(A,B),\ \mathit{append}(B,S,T),\ \mathit{derive}(T,W)\\[.5mm]

\Nex{4.}\makebox[60mm][l]{$\mathit{nonterminal}(s)\leftarrow$} 
\makebox[8mm][r]{5.~}\mathit{nonterminal}(x)\leftarrow \\[.5mm]

\Nex{6.}\makebox[60mm][l]{$\mathit{terminal}(a)\leftarrow$} 
\makebox[8mm][r]{7.~}\mathit{terminal}(b)\leftarrow\\[.5mm]

\Nex{8.}\makebox[60mm][l]{$\mathit{production}(s,[a,x,b])\leftarrow$} 
\makebox[8mm][r]{9.~}\mathit{production}(x,[\,])\leftarrow \\[.5mm]

\Nex{10.}\makebox[60mm][l]{$\mathit{production}(x,[a,x])\leftarrow$}
\makebox[8mm][r]{11.~}\mathit{production}(x,[a,b,x])\leftarrow \\[.5mm]


\Nex{12.}\makebox[60mm][l]{$\mathit{word}([\,])\leftarrow$}
\makebox[8mm][r]{13.~}\mathit{word}([A|W])\leftarrow \mathit{terminal}(A),\, 
\mathit{word}(W)\\[.5mm]

\Nex{14.}\makebox[60mm][l]{$\mathit{append}([\,],\!\mathit{Ys},\!\mathit{Ys})\leftarrow$}
\makebox[8mm][r]{15.~}\mathit{append}([A|\mathit{Xs}],
\!\!\mathit{Ys},\![A|\mathit{Zs}])\!\leftarrow \!\mathit{append}(\mathit{Xs},\!\!\mathit{Ys},\!\mathit{Zs})
\eeq
\vspace{2mm}

\noindent The relation $\mathit{derive}([s],W)$ holds iff the
word $W$ can be derived from the {start symbol} $s$
using the following productions (see clauses 8--11):

$s\rightarrow a\, x\, b$ \hspace{20mm}$x\rightarrow \varepsilon \mid a\, x \mid a\, b\, x$

\noindent The nonterminal symbols are $s$ and $x$ (see clauses~4 
and 5), the terminal symbols are $a$ and $b$ (see clauses~6 
and 7), words in $\{a,b\}^{*}$ are represented
as lists of $a$'s and $b$'s, and 
the empty word~$\varepsilon$ is represented as the
empty list $[\,]$.

The relation $\mathit{derive}(L,W)$ holds iff $L$
is a sequence of terminal or nonterminal symbols from which the word
$W$ can be derived by using the productions.

We would like to derive an efficient program for an initial goal $G$
of the form: 

\vspace{1mm}
$\mathit{word}(W),\ \neg \mathit{derive}([s],W)$

\vspace{1mm}
\noindent
which holds in the perfect model of the program 
{\it{CFParser\/}} iff $\,W$ is a word which
is \emph{not} derivable from $s$ by using the given context-free grammar. 
We perform
our two step program derivation presented in~\cite[Section~2.3]{PeP02a}.
In the first step, from goal $G$ we derive the following two clauses:


\smallskip{}
16. ~$g(W)\leftarrow \mathit{word}(W),\, \neg \mathit{new}1(W)$\nopagebreak

\smallskip
17. ~$\mathit{new}1(W)\leftarrow \mathit{derive}([s],W)$


\noindent 
In the second step, we apply the 
{\it unfold-definition-folding strategy} presented in~\cite{PrP91b}. 
We will not recall here the formal definition of this strategy. 
It will be enough to say that it is 
similar to the loop absorption strategy
we have seen in action in the above derivation starting from the
{\it Comsub}$1$ program.

For our {\it CFParser} program, at the end of the second step, we  get:

\vspace{2mm}
\makebox[100mm][l]{\makebox[38mm][l]{$g([\,])\leftarrow $}
$g([a|A])\leftarrow \mathit{new}2(A)$}
$g([b|A])\leftarrow \mathit{new}3(A)$\vspace{1mm}

\makebox[100mm][l]{\makebox[38mm][l]{$\mathit{new}2([\,])\leftarrow $}
$\mathit{new}2([a|A])\leftarrow \mathit{new}4(A)$}
$\mathit{new}2([b|A])\leftarrow \mathit{new}5(A)$\vspace{1mm}

\makebox[100mm][l]{\makebox[38mm][l]{$\mathit{new}3([\,])\leftarrow $}
$\mathit{new}3([a|A])\leftarrow \mathit{new}3(A)$}
$\mathit{new}3([b|A])\leftarrow \mathit{new}3(A)$\vspace{1mm}

\makebox[100mm][l]{\makebox[38mm][l]{$\mathit{new}4([\,])\leftarrow $}
$\mathit{new}4([a|A])\leftarrow \mathit{new}4(A)$}
$\mathit{new}4([b|A])\leftarrow \mathit{new}6(A)$\vspace{1mm}

\makebox[70mm][l]{$\mathit{new}5([a|A])\leftarrow \mathit{new}3(A)$}
$\mathit{new}5([b|A])\leftarrow \mathit{new}3(A)$\vspace{1mm}

\makebox[70mm][l]{$\mathit{new}6([a|A])\leftarrow \mathit{new}4(A)$}
$\mathit{new}6([b|A])\leftarrow \mathit{new}5(A)$

\begin{figure}[ht!]
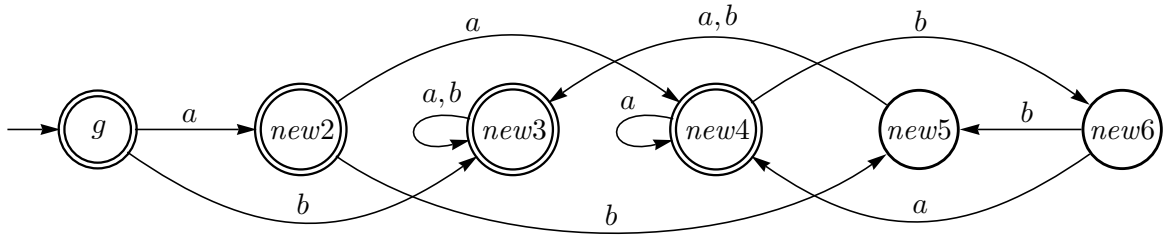

\begin{center}
\VCDraw{%
\begin{VCPicture}{(-4.4,-2.3)(14.8,3.2)} 
\SetEdgeArrowWidth{7pt}
\SetEdgeArrowLengthCoef{1.8}
\ChgStateLineWidth{1}

\FixStateDiameter{18mm}
\State[\mathit{new}5]{(12.4,0)}{new5}
\State[\mathit{new}6]{(16.9,0)}{new6}

\FixStateLineDouble{0.8}{1.2}
\StateLineDouble
\FixStateDiameter{16mm}
\State[g]{(-5.8,0)}{g}
\FixStateDiameter{19mm}     
\State[\mathit{new}2]{(-1.3,0)}{new2}
\State[\mathit{new}3]{(3.4,0)}{new3}
\State[\mathit{new}4]{(7.9,0)}{new4}

\Point{(-7.8,0)}{i}
\EdgeL{i}{g}{}

\SetArcCurvature{0.9}
\SetArcAngle{35}
\SetLoopAngle{14}
\EdgeL{g}{new2}{a}   
\LoopW{new3}{}\LabelR[.15]{a,b}
\LoopW{new4}{a}          \ArcL{new4}{new6}{}{}\LabelL[.5]{b}
\ArcR{new5}{new3}{}\LabelR[.5]{a,b}

\EdgeR{new6}{new5}{b}

\SetArcCurvature{.9}
\ArcL{new2}{new4}{a}     
\ArcR{g}{new3}{}\LabelL[.5]{b}   
\ArcL{new6}{new4}{}\LabelR[.5]{a}   

\SetArcCurvature{.65}
\ArcR{new2}{new5}{}\LabelL[.5]{b}

\end{VCPicture}
}
\end{center} \normalsize \caption{The finite automaton which accepts the words which are
{\it not} generated from $s$ by the productions: $s\rightarrow a\, x\, b$ ~and~ 
$x\rightarrow \varepsilon \mid a\, x \mid a\, b\, x$. State~$g$ is the initial state and the 
final states have double circles.\label{fig:fa}}
\end{figure}

\vspace{2mm}
\noindent This program corresponds to the deterministic finite
automaton of Figure~\ref{fig:fa}. 
Each predicate of the derived program is a state, (ii)~$g$ is the initial state, 
(iii)~a state~$p$ is final iff it has a clause of the 
form $p([\,])\leftarrow $, 
(iv)~a clause of the form $p([\ell|A])\!\leftarrow\! q(A)$
denotes a transition with label $\ell$ from~$p$ to~$q$. Note that 
the derivation of the final
program that corresponds to a finite automaton has been possible because 
the context-free grammar indeed generates a regular language.

\smallskip
Finally we present an example on how to use the transformation methodology for the
verification of program properties. 
This example is the so called {\it Yale Shooting Problem} which
is often used in temporal reasoning. 
This problem can be described and formalized as follows.

We have a person and a gun. Three \emph{events} are possible: 
(e1)~a \emph{load} event, when the gun
is loaded, (e2)~a \emph{shoot} event, when the gun shoots, and
(e3)~a \emph{wait} event, when nothing happens (see clauses 1--3 below). 
A \emph{situation} is (the result of) a sequence of events. 
A sequence is represented as a list. We assume that, as time progresses, 
the list grows `to the left', that is, given the current list~$S$ of events,
when a new event $E$ occurs, the new list of events is $[E|S]$.
In any situation, at least one of the following three
facts \emph{holds}: (f1) the person is \emph{alive}, (f2) the person
is \emph{dead}, and (f3) the gun is \emph{loaded} (see clauses 4--6 below). 

We also assume the following hypotheses (see clauses 7--11 and note 
the presence of a negated atom in clause~11).
\noindent (s1)~In the initial situation denoted by the empty
list, the person is \emph{alive}.
\noindent (s2)~After a \emph{load} event the gun is \emph{loaded.}
\noindent (s3)~If the gun is \emph{loaded}, then after a \emph{shoot}
event the person is \emph{dead}.
\noindent (s4)~If the gun is \emph{loaded}, then it is \emph{abnormal}
that after a \emph{shoot} event the person is \emph{alive}.
\noindent (s5)~\emph{Inertia Axiom\/}: If a fact $F$ holds in a situation $S$ and
it is not abnormal that $F$ holds after the event $E$ following
$S$, then $F$ holds also after the event $E$. \smallskip

The following locally stratified program \!{\it YSP\/} formalizes
the above statements. A similar formalization is in a paper by Apt and 
Bezem~\cite{ApB91}.

\vspace{2mm}

\noindent
\makebox[110mm][l]{~~\makebox[55mm][l]{~1.~$\mathit{event}(\mathit{load})\leftarrow $}
~2.~$\mathit{event}(\mathit{shoot})\leftarrow $}~3.~$\mathit{event}(\mathit{wait})\leftarrow $ \vspace{1mm}

\noindent
\makebox[110mm][l]{~~\makebox[55mm][l]{~4.~$\mathit{fact}(\mathit{alive})\leftarrow $}
~5.~$\mathit{fact}(\mathit{dead})\leftarrow $}~6.~$\mathit{fact}(\mathit{loaded})\leftarrow $\vspace{1mm}

\noindent
\makebox[55mm][l]{~~~7.~$\mathit{holds}(\mathit{alive},[\,])\leftarrow $}
~~~8.~$\mathit{holds}(\mathit{loaded},[\mathit{load}|S])\leftarrow $\vspace{1mm}

\noindent
~~~9.~$\mathit{holds}(\mathit{dead},[\mathit{shoot}|S])\leftarrow \mathit{holds}(\mathit{loaded},S)$\vspace{1mm}

\noindent
~~10.~$ab(\mathit{alive},\mathit{shoot},S)\leftarrow \mathit{holds}(\mathit{loaded},S)$\vspace{1mm}

\noindent
~~11.~$\mathit{holds}(F,[E|S])\leftarrow \mathit{fact}(F),\  \mathit{event}(E),\  \mathit{holds}(F,S),\  \neg\, ab(F,E,S)$\vspace{1mm}

\noindent
\makebox[55mm][l]{~~12.~$\mathit{append}([\,],\!\mathit{Ys},\!\mathit{Ys})\leftarrow$}
~~13.~$\mathit{append}([A|\mathit{Xs}],\!\mathit{Ys}, [A|\mathit{Zs}]) \leftarrow  \mathit{append}(\mathit{Xs},\! \mathit{Ys}, \mathit{Zs})$

\vspace{2mm}

\noindent By applying SLDNF-resolution~\cite{Llo87}, Apt and Bezem showed that 
$\mathit{holds}(\mathit{dead},[\mathit{shoot},\mathit{wait},\mathit{load}])$ is true
in the perfect model of program \!{\it YSP\/}.
Now we consider a property $\Gamma$ which {\it cannot\/} be shown by SLDNF-resolution 
(see~\cite{PeP02a}):


\vspace{1mm}
\noindent ${\Gamma}\ \equiv\ \forall S\, 
(\mathit{holds}(\mathit{dead},S)$
$~\rightarrow~~ \exists S0,S1,S2,S'\, (\mathit{append}(S2,[\mathit{shoot}|S1],S'),\, \mathit{append}(S',[\mathit{load}|S0],S)))$

\vspace{1mm}
\noindent Property~$\Gamma$ means that the fact that the person is 
\emph{dead\/} in the current
situation $S$ implies that in the past there was a \emph{load} event followed, 
possibly not immediately, 
by a \emph{shoot} event. Thus, since time progresses `to the left', $S$ is a list
of events of the form:
 $[\ldots, \mathit{shoot}, \ldots, \mathit{load}, \ldots]$.

In the first step of our two step verification method (see~\cite[Section~2.3]{PeP02a}), we
apply the Lloyd-Topor transformation~\cite[page 113]{Llo87}
starting from the statement: $g\leftarrow {\Gamma}$ (where $g$ is a new predicate name) and we derive the following clauses:

\vspace{1mm}
\noindent
~~14. $g\leftarrow \neg \mathit{new}1$\vspace{1mm}

\noindent
~~15. $\mathit{new}1\leftarrow \mathit{holds}(\mathit{dead},S),\  \neg \mathit{new}2(S)$\vspace{1mm}

 \noindent
~~16. $\mathit{new}2(S)\leftarrow \mathit{append}(S2,[\mathit{shoot}|S1],S'),\ 
 \mathit{append}(S',[\mathit{load}|S0],S)$

\vspace{2mm}

\noindent At the end of the second step, 
after a few iterations of the {\rm unfold-definition-folding strategy}
and after the deletion of all definitions of predicates which 
are not required by $g$, we are left with the single clause: $g\leftarrow $. 
Details can be found in~\cite{PeP02a}. 

Since $g$ holds in the (perfect model of the) final program, we have that 
property~${\Gamma}$ holds in the (perfect model of the)
final program. Thus, ${\Gamma}$ holds also in the initial program made out
of clauses~{1--13}.


\smallskip

Much more recently we have explored some verification techniques based on the 
transformation of {\it constrained Horn clauses,} also in the case of imperative and 
functional programs~\cite{De&17b} and in the case of business processes (see, for 
instance,~\cite{De&16b}). 
This recent work has been done in cooperation with
Emanuele De Angelis and Fabio Fioravanti. They also have been working and 
still work in the implementation and development of an automatic
transformation and verification tool~\cite{De&14b}, which was originally set 
up by Ornella Aioni and Maurizio Proietti.

\section{Future Developments} 
\label{sec:FutureDevelopments}
Reviewing my research activity when writing this paper, 
I realized that many topics and issues
would need a more accurate analysis and study. It would be difficult to
list them all, but I have been encouraged to mention at least 
some of them. I hope that these suggestions may be useful for researchers in 
the field and they may find these suggestions of some interest.

Concerning the theory of combinators and WCL presented in 
Section~\ref{sec:SignalTheory-CombinatoryLogic}, one should note that the combinator
$X \equiv B(B(BB)B)(BB)$ we have presented has
parentheses and one could consider to construct a $B$-combinator, 
call it $\widetilde B$,
which places those parentheses in a sequence of seven $B$'s, so that
$\widetilde B BBBBBBB >^{*} B(B(BB)B)(BB)$. A routine construction, 
following~\cite{BaP75}, shows that
$\widetilde B$ is, in fact, $B(B(B(BB)B)B)(BB)$. The relation between combinators
 $X$ and $\widetilde B$ could be 
for the reader a stimulus for studying the process of placing parentheses in a list
of variables, that is, the process of 
constructing a binary tree from the list of its leaves. 

One can start by 
considering, first, the use of {\rm regular} combinators only. 
A combinator~$X$ is said to be {\it regular\/}
if its reduction is of the form $Xx_{1}\ldots x_{n} > x_{1}t_{2}\ldots t_{m}$, 
where $t_{2},\ldots, t_{m}$ are terms made out of $x_{2}, \ldots,x_{n}$ only. 
A particular regular combinator for placing parentheses is, indeed, $B$.
Similarly, one could study the permutative and duplicative 
properties of the regular combinators~$C$ (defined by $Cxyz > xzy$) and~$W$ 
(defined by $Wxy > xyy$) and other regular (or non-regular) combinators. 
This study will improve the results reported 
in the classical book by Curry and Feys~\cite[Chapter~5]{CuF74}.

For Section~\ref{sec:Finite_InfiniteComputation} one could develop the
techniques presented in~\cite{Pet80a}. Those developments can be useful in the 
area of Term Rewriting Systems for constructing
terms with infinite behaviour.

For the issues considered in Section~\ref{sec:ProTrans} on 
Program Transformation,
it will be important to investigate how to invent the multiplication 
operation, having at our disposal in the initial program version
only the addition operation. Generalizations of 
various kinds can be suggested as we have done in this paper, 
but an interesting technique would be the one based on the idea of deriving 
multiplication as the {\it iteration} of additions. Then, in an analogous
way, exponentiation can be invented as the iteration of multiplications,
thus allowing us to derive even more efficient 
programs.~The idea of iteration can hopefully be
generated by mechanically analyzing the m-dags constructed by unfolding and
looking at repeated patterns. 

For Sections~\ref{sec:TuplingListIntro} and~\ref{sec:LA}, 
it could be important to mechanize the techniques we 
have presented there,
and in particular those for finding the suitable tuples of functions 
and suitable lambda-abstractions via the analysis of: (i)~cuts and pebble 
games in the m-dags, and (ii)~subexpression mismatchings,
respectively.

For Section~\ref{sec:CommPar} one can provide an implementation of the functional 
language with communications we have proposed so that one can execute programs 
written in that language.
One may also: (i)~automate the process of adding  
communications to functional programs
for improving their efficiency by making use of the properties of the 
functions to be evaluated, and (ii)~automate the reasoning on the modal 
theories presented in~\cite{PeS83} in which one can prove correctness
of those communications. Thus, one will have a machine-checked 
proof of correctness of the communications which have been added.

For Section~\ref{sec:TransfVerifinLogic} a possible project is
to construct a transformation system of logic programs with goals 
as arguments in which: (i)~one can run the programs according 
to the operational semantics we have defined in our paper~\cite{PeP99b}, and
(ii)~one can apply the various transformation rules (definition introduction, 
unfolding, folding, goal replacement) we have listed in that paper.

\section{Acknowledgements} 
\label{sec:Ack}
My gratitude goes to the various people who taught me Computer Science and among
them, Paolo Ercoli, Rod Burstall, Robin Milner, Gordon Plotkin, John Reynolds, 
Alan Robinson, and Leslie Valiant.
I am also grateful to my colleagues and students. In particular, 
I would like to thank
Anna Labella, Andrzej Skowron, Maurizio Proietti, Valerio Senni,
Sophie Renault, Fabio Fioravanti, and 
Emanuele De Angelis.
A very special thank goes to Maurizio, who for many years has been 
`so devoted to me, 
so patient, so zealous', 
as John Henry Newman said of his friend Ambrose St.~John~\cite[page~190]
{New01}. Maurizio
has been for me an unvaluable source of inspiration and strength.

Many thanks to Andrei Nemytykh for inviting me to write this paper and for his comments, 
and Maurizio Proietti and Laurent Fribourg for their friendly help and encouragement. 


\bibliographystyle{eptcs}

\end{document}